# Computational Analysis for the Rational Design of Anti-Amyloid Beta (A$\beta$) Antibodies

D'Artagnan Greene[1], Theodora Po[1], Jennifer Pan[1], Tanya Tabibian[1], and Ray Luo[1,2,,3,4]

1. Department of Molecular Biology and Biochemistry,
2. Chemical and Materials Physics Graduate Program,
3. Department of Biomedical Engineering, and
4. Department of Chemical Engineering and Materials Science,
University of California, Irvine, CA 92697

Please send correspondence to: rluo@uci.edu

## Abstract

Alzheimer's Disease (AD) is a neurodegenerative disorder that lacks effective treatment options. Anti-amyloid beta (A$\beta$) antibodies are the leading drug candidates to treat AD, but the results of clinical trials have been disappointing. Introducing rational mutations into anti-A$\beta$ antibodies to increase their effectiveness is a way forward, but the path to take is unclear. In this study, we demonstrate the use of computational fragment-based docking and MMPBSA binding free energy calculations in the analysis of anti-A$\beta$ antibodies for rational drug design efforts. Our fragment-based docking method successfully predicted the emergence of the common EFRH epitope, MD simulations coupled with MMPBSA binding free energy calculations were used to analyze scenarios described in prior studies, and we introduced rational mutations into PFA1 to improve its calculated binding affinity towards the pE3-A$\beta_{3-8}$ form of A$\beta$. Two out of four proposed mutations stabilized binding. Our study demonstrates that a computational approach may lead to an improved drug candidate for AD in the future.

# Introduction

Alzheimer's Disease (AD) is an incurable neurodegenerative disorder that leads to steady memory and cognitive function loss, culminating in death. At present there is no cure for AD, and there is a notable absence of treatment options that can reverse or effectively slow progression of the disease. At the level of brain tissue, AD is characterized by both the appearance of extracellular, fibrous plaques that are built up from the polymerization of amyloid beta (A$\beta$) peptides[1] and the appearance of intracellular neurofibrillary tangles that consist of hyperphosphorylated tau proteins[2]. Amyloid fibril deposits are hallmarks of several neurodegenerative diseases[3], and the amyloid hypothesis states that an excessive buildup of A$\beta$ plaques in the brain is responsible for the cognitive decline observed in AD patients. It suggests that clearing A$\beta$ plaques from the brain would help inhibit or reverse progression of the disease.

The amyloid hypothesis has been the leading theory driving therapeutic approaches for the treatment of AD for over two decades[4]. The most common therapeutic approach to AD treatment is immunotherapy[5-7]. Several active and passive anti-A$\beta$ immunotherapies that target A$\beta$ species in the brain have advanced to clinical trials, but the results thus far have been disappointing. The vaccine AN1792, which targeted a full-size A$\beta$ 1-42 peptide, advanced to human clinical trials in 2001 but was terminated in the phase II trial after 6% of the treated patients developed meningoencephalitis[8].

The result of the AN1792 trial led to the development of several passive immunization approaches over the next several years. Shorter regions of the A$\beta$ sequence were used to develop monoclonal antibodies that target different cell types in the immune system. The N-terminal A$\beta_{1-15}$ sequence was used for A$\beta$ specific B cell epitopes while A$\beta_{16-42}$ was used for A$\beta$ specific T cell epitopes[9]. The choice of the epitope has a crucial effect on the ability of the antibody to bind to its amyloid target species. The N-terminal epitope is accessible to antibody binding in aggregated forms of A$\beta$ while the central and C-terminal epitopes are only able to bind to the antibody in monomeric, or perhaps small oligomeric, forms of A$\beta$ due to the central and C-terminal epitopes being inaccessible in mature fibril structures.

Initially, two anti-A$\beta$ monoclonal antibodies advanced to clinical trials targeting distinct epitopes/species of A$\beta$. Bapineuzumab primarily targets insoluble amyloid plaques via the hydrophilic N-terminal epitope of A$\beta_{1-5}$. Initial results looked promising, but during phase II of the clinical trial a serious side effect appeared as 10% of patients developed vasogenic edema[10]. Solanezumab targets soluble monomeric A$\beta$ peptides via the hydrophobic central A$\beta_{16-24}$ epitope. In clinical trials, solanezumab had a much-improved safety profile as adverse side effects such as meningoencephalitis, microhemorrhage, and vasogenic endema were not obserseved[11-12]. On the other hand, questions about solanezumab's efficacy in reducing neuritic plaque burden arose, and recently it was announced that the phase III trials for solanezumab had failed to show a significant benefit in slowing cognitive decline for mild-to-moderate AD patients[13-16].

The question of how to improve a monoclonal antibody to treat AD is not straightforward. Removing harmful side effects is one issue, but to complicate matters, A$\beta$ species exhibit a high degree of structural polymorphism, and several other A$\beta$ species have emerged as potential disease-causing agents that would presumably need to be removed in an effective AD treatment[17]. For example, normally rare, N-terminal truncated variants of A$\beta$ have been found in much higher concentrations in the stable, neurotoxic A$\beta$ aggregates that are found in severe AD cases. For example, plaques can be enriched by as much as 50% with the pE3-A$\beta$ form of A$\beta$[18-19], and pE3-A$\beta$ has become a target in antibody development[20-22].

The existence of such polymorphic amyloid targets implies that targeting a single epitope associated with a single A$\beta$ species might not be enough for an antibody to treat AD effectively. At the moment, next generation monoclonal antibodies that can bind to multiple A$\beta$ epitopes and species are currently in clinical trials. Gantenerumab binds at nanomolar affinity to several A$\beta$ species (with a $K_D$ of 0.6 nM for A$\beta$ fibrils, 1.2 nM for A$\beta$ oligomers, and 17 nM for A$\beta$ monomers), and it recognizes two epitopes within A$\beta$: the N-terminal EFRHDSGYEV sequence and a central region from the sequence VFFAEDVGSN[23]. Similarly, crenezumab, despite being generated by immunization with the N-terminal A$\beta_{1-16}$ epitope[24-25], has been shown to bind to monomeric and oligomeric forms of A$\beta$ via the central A$\beta$ epitope[7, 25-26]. Due to similarities it shares with solanezumab, the ability to bind to the central epitope of A$\beta$ has been emphasized in studies of crenezumab[27], and a co-crystal structure of crenezumab (more specifically, CreneFab) was recently obtained bound to an A$\beta$ peptide containing the central epitope[26]. Nevertheless, crenezumab has also been shown to bind to amyloid fibril species[25-26], which is puzzling since the central epitope is not readily accessible for binding in mature fibril structures. Recently, aducanumab was heralded as possibly the first "successful" anti-A$\beta$ antibody as it was able to clear A$\beta$ plaques thoroughly at the highest dosage and was shown to reduce cognitive decline in an early Phase III clinical trial that took place over the course of a year[28]. Aducanumab has been reported to bind to both N-terminal and central epitopes of A$\beta$, accounting for binding to both fibril and oligomeric forms of A$\beta$[29].

In this study, we have taken a computational approach to explore how an anti-A$\beta$ antibody may bind to one or more A$\beta$ epitopes in its antigen-combining site and to demonstrate how rational mutations aimed at modifying the antigen-combining site can be of use in rational drug design efforts. Using computational techniques, we explored several open questions from the literature including: 1) which residues in A$\beta_{1-42}$ are most important in the initial binding event that anchors A$\beta$ to a given antibody structure?, 2) how might antibodies like gantenerumab, crenezumab, and aducanumab bind to both the hydrophillic N-terminal and hydrophobic central epitopes?, and 3) can we predict a useful mutation that improves the binding affinity of a polymorphic form of A$\beta$ towards an anti-A$\beta$ antibody?

To address the first question, we employed an unbiased, fragment-based docking method to probe the antigen-combining site of various anti-A$\beta$ antibodies using single amino acid residues. To address the second question, we ran molecular dynamics (MD) simulations using the available crystal structures of gantenerumab and crenezumab bound to short A$\beta$ peptides[23, 26] (a crystal structure of aducanumab was not available for us to analyze at this time). Using Amber 16[30], we carried out Molecular Mechanics Poisson Boltzmann Surface Area (MMPBSA) binding free energy calculations to help us explore the possibility that both N-terminal and central A$\beta$ epitopes are recognized by each antibody. For the third question, we built on the study of PFA1 bound to A$\beta_{2-7}$ and pE3-A$\beta_{3-8}$ that was carried out previously by Gardberg et al.[31] Improving the binding affinity between pE3-A$\beta_{3-8}$ and PFA1 is an accessible test case given that the sequence, charge characteristics, and binding pose for pE3-A$\beta_{3-8}$ are similar to the wildtype. A computational approach that provides a way to probe the interactions between an Aβ peptide and an anti-Aβ antibody, and acts as a means to test the efficacy of changes to the antigen-combining site, may be key to producing the most effective drug candidate for this disease in the long term.

## Materials and Methods

### Computational docking of amino acid residues to anti-Aβ antibodies

We carried out an unbiased, fragment-based computational docking study to examine the initial binding characteristics of the antigen-combining site on various anti-Aβ antibodies. 16 amino acids were individually docked to each antibody, comprising the full Aβ 1-42 monomer sequence. Each single amino acid residue was generated using the sequence command in xleap from Amber 16[30]. Three types of amino acid residue fragments were initially tested using our docking protocol: 1) the default amino acid residue that contained charged N- and C-terminal groups on the backbone, 2) a neutralized version where methyl groups were attached to both the N- and C-terminal groups to remove the backbone charges, and 3) a non-physiological fragment where two hydrogens on the N-terminus and an oxygen on the C-terminus were omitted from the structure. We compared the results of the docking for a few test residues using each method above to the actual binding sites observed in the PFA1 and PFA2 crystal structures and also to the leading ligand-free hotspots found by submitting the PFA1 and PFA2 apo structures to the FTMap server[32-33]. Of the three options, we found that method 3 worked the best. For method 1, the zwitterionic terminal backbone charges were capable of binding to antibody hot spots in place of the side chain functional groups, and while method 2 neutralized the terminal backbone charges, we found that some fragments were unable to bind to certain hot spots due to steric issues brought about by the attached methyl groups. Method 3 removed both the terminal charges and the steric issue of adding a methyl group to the backbone. Therefore, residues were generated using method 3 for our full analysis. We should note that there are other possible ways to generate single amino acid fragments that were implemented in other computational studies, such as removing the backbone entirely and replacing the alpha carbon with a proton[34]. In order to obtain a single amino acid fragment that behaves more like an amino acid in the middle of a longer peptide sequence as opposed to a terminal amino acid, the side chain for each amino acid should be unaltered in the fragment while the effect of the backbone should be minimized in some way to prevent it from interfering or competing with the binding of the side chain.

After obtaining an amino acid residue, Open Babel[35] was used to convert pdb files to pdbqt files for both the antibody and the residue, and then the residue was docked to the antibody using Autodock Vina/SMINA[36]. The residue was allowed to search the entire antibody surface within a 100 $Å^3$ box that was centered on the antigen-combining site with a search exhaustiveness of 128. Using the default settings in Autodock Vina/SMINA, the top nine docked results for each residue, ranked by their most stable binding free energy values, were subsequently used for our analysis.

To carry out a comparison study with a variety of anti-amyloid antibodies, this method was performed on the previously published crystal structures of bapineuzumab, solanezumab, gantenerumab, crenezumab, ponezumab, PFA1, and PFA2 in both the holo and apo forms (if available) of each structure (PDB IDs: 4OJF, 4XXD, 5CSZ, 5VZX, 5VZY, 3UOT, 2IPU, 2IPT, 2R0W, and 2IQ9). The PDB file for each antibody complex was edited prior to docking to remove everything except for the residues of the isolated antibody structure.

**Molecular dynamics simulations**

To prepare the various complexes that were used in this study for MD simulations, the program Modeller[37] was used to accomplish two purposes: 1) to model in any missing residues that were not present in the original PDB file, and 2) if applicable, to generate a homology model containing a new A$\beta$ peptide that was not present in the original PDB file. Prior to using Modeller, the PDB file was processed to remove everything except for the A$\beta$ peptide-antibody complex. After using Modeller, the structure was further processed using the program xleap in Amber 16 to add in hydrogens, water, counter ions, and disulfide bonds. A TIP3P water box was constructed using the solvatebox command to ensure that all atoms in the starting structure were no less than 12 Å from the edge of the water box. Counter ions (i.e. chloride ions) were added to the solvated system to give a total net charge of zero. Disulfide bonds were added between pairs of residues that were indicated as having disulfide bonds in the accompanying PDB files. For the pE3-A$\beta_{3-8}$ model (PDB ID: 3EYS), antechamber was used to parameterize the pyroglutamate (PCA) moiety.

For each MD run, a 1000 step minimization was carried out with 500 steps of steepest descent followed by 500 steps of conjugate gradient using a non-bonded cutoff of 8.0 Å. The system was then heated up to a constant 300 K over a period of 50 ps under NVT conditions employing the Langevin thermostat. The density was equilibrated over an additional 50 ps under NTP conditions, and an equilibration under NVT conditions was undertaken for approximately 250 ns. Finally, a 50 ns production run was carried out to bring the total simulation time to 300 ns. Due to the size and high conformational flexibility of both the amyloid peptide and the antibody, and the extensive use of homology modeling, a long equilibration/production run was needed in order to achieve acceptable convergence for our MMPBSA binding free energy calculation (see **MD method validation** in Results and Discussion). Afterwards, the MD trajectory of our production run was visualized using UCSF Chimera[38]. In addition, to characterize alterations in binding patterns in the antigen-combining sites of our peptide-antibody structures, residue-to-residue percent occupancy calculations were carried out between select residues in the A$\beta$ peptide and on the antibody surface. These calculations used a cutoff distance of 10 Å over 5000 frames collected at equal intervals from the 50 ns production run. This relatively high cutoff distance was chosen due to the high conformational flexibility of certain ligands in the antigen-combing site which led to a difficulty in identifying specific binding interactions; as a result, we do not use this measure to indicate such binding interactions per say, but rather, a low percent occupancy indicates a major decrease in the likelihood that the two residues in question can participate in such a binding interaction at a given site.

**MMPBSA binding free energy calculations**

5000 frames, taken at equal intervals over the 50 ns production run, were used to calculate the MMPBSA binding free energy[39-44] for each A$\beta$ peptide-antibody complex. All PBSA calculations were carried out using the PBSA program[45-53] in the AMBER 16 package.[30, 54] For this calculation, inp = 2 was used for the non-polar solvation model,[55-56] radiopt=0 was used for the intrinsic atomic radii, and the ionic strength was set to 100 mM. All other settings were kept at the default values used in Amber 16.[30, 54] Due to the rather high uncertainties in normal mode analysis, the entropy contribution was neglected in our binding affinity calculations. Experimental binding affinities were compared to our calculated binding affinities by converting $K_D$ values into binding free energies using:

$$\Delta G = RT\ln(K_D)$$

where R = 1.987 * 10$^{-3}$ kcal mol$^{-1}$ K$^{-1}$, T = 300K, and $K_D$ is the dissociation constant in units of M.

## Results

### Computational docking of amino acid residues to anti-A$\beta$ antibodies

To address the question of which amino acids are key to the initial binding and anchoring of an A$\beta$ peptide to an anti-amyloid antibody, we carried out an unbiased fragment-based computational docking search using all 16 unique amino acid residues that appear in the A$\beta_{1-42}$ sequence. To minimize conformational effects from the polypeptide ligand, we chose to dock single amino acid residue fragments to the antibody surface. The goal in carrying out this step is to determine which single amino acid(s) are most important to the initial binding event. Previous studies have indicated that antibodies with extended epitopes that exhibit promiscuous binding have a tendency to rely on a small number of key semi-conserved anchor residues when binding[57-58]. The ability to identify key anchor residue(s) provides an important clue towards understanding how certain anti-A$\beta$ antibodies can initially recognize both the N-terminal and central A$\beta$ epitopes despite the low sequence similarity between the two regions. Two criteria from the output were considered to be relevant in determining the likelihood that an amino acid will bind with high affinity to the antigen-combining site of the antibody: 1) which residues have the strongest binding interaction at the antigen-combining site?, and 2) which residues find the antigen-combining site most consistently?

**TABLE 1** lists the most stable binding affinities for the top 10 ranked amino acid residues bound to the antigen-combining sites of ten anti-A$\beta$ antibody structures. From **TABLE 1**, we observe that approximately 95% of the amino acid residues that are ranked in the top 5 appear in the A$\beta$ 1-23 region while 13% of the amino acid residues that are ranked in the top 5 appear in the A$\beta$ 24-42 region. Furthermore, 57% of the amino acid residues that are ranked in the top 5 are found in the N-terminal DAEFRH epitope while 39% appear in the central KLVFFAED epitope. In particular, nine out of the ten antibodies have at least three out of four residues in the EFRH sequence appear in their top 5 ranking; the only exception is gantenerumab where E and H are tied for sixth place. These results are consistent with the noted tendency of the N-terminal DAEFRH epitope to bind to most anti-A$\beta$ antibodies[59].

In **TABLE 2**, we examine the number of docked amino acids found at the antigen-combining site for each residue across the 10 different anti-A$\beta$ antibodies. If we consider only the amino acids that were able to dock at the antigen-combining site more than half of the time, 79% appear in the A$\beta$ 1-23 region while 47% appear in the A$\beta$ 24-42 region. Of the residues that found the correct binding site more than half the time, 39% are found in the N-terminal DAEFRH epitope while 46% appear in the central KLVFFAED epitope. In contrast to **TABLE 1**, which displayed similar results in the residue ranking for each antibody, **TABLE 2** displayed clear differences between the different highly ranked residues docked to different antibodies and even showed some noticeably large differences for the holo and apo forms of the same antibody. These latter differences are presumably due to differences in the specific conformation of the antibody in the holo and apo crystal structures. It is worth pointing out that while charged and aromatic residues appear to dominate the top ranks in **TABLE 1**, the ability of fragments to find the active site, as observed in **TABLE 2**, does not correlate as strongly with polarity as many polar and non-polar residues tend to locate the antigen-combining site with similar ease.

An examination of **TABLES 1 and 2** reveals an important trend in the binding pattern that appears to be consistent across the various antibodies that we have studied; the top two residues on A$\beta$ 1-42 that have the most stable binding free energy are either phenylalanine or tyrosine in all cases, and both residues are also near the top of the residues that are consistently docked to the antigen-combining site. Since phenylalanine and tyrosine are structurally identical except for a hydroxyl group, the strong binding free energy points to the presence of an aromatic binding pocket on the antibody that is important for binding A$\beta$. The existence of such a binding pocket has been pointed out before as a key binding site for phenylalanine by several anti-A$\beta$ antibody crystallographers[23, 26, 59-60]. In addition, the two prominent A$\beta$ epitopes, the N-terminal DAEFRH epitope and the central KLVFFAED epitope, correspond to the only two locations in the A$\beta$ 1-42 sequence where phenylalanine appears. Our docking results, taken together with these observations, point towards phenylalanine as a leading candidate for the most important anchor residue in the initial binding of A$\beta$ epitopes to an anti-amyloid antibody.

**MD method validation**

Although the computational docking of single amino acid residues may give us some insight as to which amino acids might bind first to the antigen-combining site of an anti-A$\beta$ antibody, it does not necessarily help us understand how binding affinity emerges for an extended polypeptide. After the first residue in a polypeptide binds to the antigen-combining site, a previously accessible, high affinity binding site now becomes unavailable to other residues, and the binding of the first residue also restricts the search space where other residues can bind. In addition, our docking protocol lacks many key factors, such as the presence of water and the conformational motion of the full polypeptide and antibody, which have both been identified as being important factors in determining the selectivity of A$\beta$ peptide-antibody binding[59].

Therefore, to study the binding affinity of extended A$\beta$ peptides to amyloid antibodies, we turned to MD simulations and MMPBSA binding free energy calculations using the Amber 16 software suite[30]. First, we needed to validate our computational approach using previously published experimental data. To do this, we compared our MMPBSA binding free energy calculations to the full set of experimental binding affinities reported in the study carried out by Gardberg et al. on PFA1 and PFA2 bound to various A$\beta$ peptides[60]. **FIGURE 1** shows the correlation of our MMPBSA calculations with the experimental data (the numerical data are given in **TABLE 3)**. With a Pearson's R value of 0.95, our MMPBSA calculations show very good agreement with the trend seen in the experimental binding affinity data. **FIGURE S1** shows the convergence of our MMPBSA free energy values taken over the entire 50 ns production run. It is seen that our data set shows reasonable convergence over this time frame. One data point, for the A$\beta_{1-8}$ peptide bound to PFA1, converged very slowly and underwent a substantial change in its binding affinity over the 50 ns MMPBSA calculation. To verify that the MMPBSA result for A$\beta_{1-8}$ bound to PFA1 had converged properly, we collected 10 ns of additional MD simulation data and ran a 60 ns MMPBSA calculation using the simulation data from 250 ns to 310 ns. The 60 ns MMPBSA result showed that the 50 ns MMPBSA result had indeed already converged (see **FIGURE S2**). In general, we found that running MD simulations for a total simulation time of 300 ns, and using 5000 frames taken from the last 50 ns for MMPBSA calculations, was sufficient to produce acceptable convergence for our data. This became the standard protocol that we used for any subsequent analysis.

## The importance of phenylalanine to the stable binding of A$\beta_{2-7}$ to PFA1

To test the importance of phenylalanine to A$\beta$ peptide-antibody binding, we studied an experimental scenario discussed by Gardberg et al.[60]. The authors carried out a binding assay for A$\beta_{2-7}$ and several other A$\beta_{2-7}$ sequence variants bound to PFA1. They demonstrated that binding affinity was lowered (from 60 nM to 3400 nM), but not completely abolished, when the wild type sequence, AEFRHD, was changed to the human glutamate receptor interacting protein 1 (or Grip1) sequence, AKFRHD. This was surprising since the charge characteristics completely changed from the negative glutamate residue to the positive lysine residue in the Grip1 peptide[59]. On the other hand, no binding to PFA1 was observed at all when AEFRHD was mutated to AEIRHD (the Position 4 or Pos4 mutant) despite the swapping of two non-polar hydrophobic residues[59-60]. We constructed homology models for the Grip1 and Pos4 mutants, carried out MD simulations, calculated the MMPBSA binding free energies for each, and compared the results to the A$\beta_{2-7}$ MMPBSA binding free energy from our method validation. **TABLE 4** shows that the MMPBSA binding affinities qualitatively reproduce the experimental results from the Gardberg study quite well.

To examine why the Pos4 (AEIRHD) mutation is much more severe than the Grip1 (AKFRHD) mutation, we first constructed RMSD plots of the 50 ns production runs for each of the three relevant structures. The average RMSD values for the two mutants, Grip1 and Pos4, are clearly higher than the average RMSD for the A$\beta_{2-7}$ structure indicating a larger degree of structural change from the initial structure for the two mutants (see **FIGURE S3**). However, the RMSD plots provide no indication that the Pos4 mutant has a significantly lower binding affinity than the Grip1 mutant.

Next, the MD trajectories of the three complexes were visualized and compared with one another. It was observed that both mutant residues were no longer able to bind to their original binding pockets in comparison to the wildtype A$\beta_{2-7}$ peptide. **FIGURE 2** shows a representative frame, taken from the halfway point of the MD trajectory, to illustrate the situation for each structure. In the Grip1 structure, the binding pocket and the lysine residue have separated from each other indicating that the binding contact between the two has been disrupted (see **FIGURE 2B**). However, the rest of the residues in the A$\beta_{2-7}$ sequence remain bound in their proper orientations (in agreement with structural observations made by Gardberg at al.[60]), which suggests that the loss or alteration of the single glutamate binding interaction is the major cause of the decreased binding affinity of PFA1 for Grip1.

The Pos4 mutation is more complicated. We note first that the isoleucine residue is displaced outside of the deep phenylalanine binding pocket (**FIGURE 2C**). Another noteworthy difference is that, when the isoleucine residue was pushed out of the pocket, the front half of the A$\beta$ peptide chain rotated. The glutamate residue disassociated from its normal binding pocket and was found instead binding to a nearby binding pocket that normally binds to either aspartate or alanine, which appear at positions 1 and 2 of the full A$\beta_{1-8}$ peptide respectively (**FIGURE 3**). Our unbiased computational docking data predicted that the Pos4 mutant would have a larger destabilizing effect on the binding affinity compared to the Grip1 mutant. From **TABLE 1**, phenylalanine had a top binding affinity of -5.9 kcal/mol and -6.1 kcal/mol for the holo and apo forms of PFA1 respectively, while glutamate had a top binding affinity of -4.5 kcal/mol and -4.8 kcal/mol for the holo and apo forms. However, the change in orientation of the binding pose in the Pos4 mutant for the full peptide clearly could not be captured by calculating the binding affinity for a single amino acid residue docked to the antibody surface.

## Analysis of gantenerumab and crenezumab binding to multiple A$\beta$ epitopes

Using our prior observation that phenylalanine is a very important residue for the binding of A$\beta$ peptides to anti-amyloid antibodies, we also examined the antigen-combining sites of gantenerumab and crenezumab with the aim of discovering how these two antibodies might bind to both N-terminal and central epitopes of A$\beta$ peptides. We first ran MD simulations and calculated MMPBSA binding free energies for both gantenerumab and crenezumab bound to the A$\beta$ peptides observed in their crystal structures (PDB IDs: 5CSZ and 5VZY). Afterwards, we generated homology models containing the other prominent A$\beta$ epitope that was not present in the original crystal structure for both gantenerumab and crenezumab.

Four homology models that featured the new peptide were generated where: 1) each new A$\beta$ peptide was modeled in both forward and reverse orientations across the antigen-combining site, and 2) the location of phenylalanine residues in the A$\beta$ epitopes were used for the initial alignment in the antigen-combining site. Calculated MMPBSA binding free energies are given in **TABLE 5** while the convergence for our MMPBSA values is demonstrated in **FIGURE**

**S4**. Only the most stable homology model of the transposed Aβ peptide bound to each antibody was used for analysis; we have labeled them as "gantenerumab forward" and "crenezumab reverse 2" to reflect their initial alignment. The other transposed epitope structures were less stable and were deemed unsuitable for further analysis (data not shown).

The N-terminal peptide, DAEFRHDSGYE, bound to gantenerumab yielded a very stable calculated binding free energy of -33.8 kcal/mol in comparison to the experimental value reported for the Aβ monomer bound to gantenerumab which was -10.7 kcal/mol[23]. **FIGURE 4** compares the initial pose of the wildtype N-terminal peptide bound to gantenerumab (**FIGURE 4A** and **4B**) to snapshots taken at the first, middle, and last frames of the MD production run (**FIGURE 4C**, **4D**, **4E**, and **4F**) for our most stable model of gantenerumab bound to the central Aβ peptide.

The source of the strong binding affinity for the N-terminal Aβ peptide seems apparent from an examination of **FIGURE 4B** as the antigen-combining site of gantenerumab exhibits several positive (blue) electrostatic contact sites on its surface that correlate with the many negatively charged aspartate and glutamate (red) residues that appear on the N-terminal Aβ peptide. Calculated percent occupancy values for various sites can be found in **TABLE S1**. Comparing the original N-terminal Aβ peptide binding pose in **FIGURE 4B** to the binding pose of the transposed central Aβ peptide in **FIGURE 4D**, **4E**, and **4F**, we see that most of these electrostatic contacts have been lost when the central Aβ peptide is bound as it does not contain as many charged residues as the N-terminal peptide. However, there appear to be at least two potential binding interactions when the central Aβ peptide is bound to gantenerumab: E22 to R57 and F19 to F119 (both from the VFFAED portion of the epitope) appear to be close enough to interact in both structures. The percent occupancy for E3 to R57 of 100% when the N-terminal Aβ peptide is bound only drops to 85.58% for E22 to R57 when the central Aβ peptide is bound, and the percent occupancy for F4 to F119 of 96.12% when the N-terminal Aβ peptide is bound actually increased to 99.92% for F19 to F119 when the central Aβ peptide is bound.  However, it is worth noting that D23 appears to be competing to some extent with E22 for occupancy near the R57 binding site as indicated by a noticeable 45.54% occupancy for D23 to R57 when the central Aβ peptide is bound.  In our figures this competition can even be seen, as E22 is clearly bound to the pocket in **FIGURE 4E** whereas D23 is bound to that same site in **FIGURE 4F**. Outside of these two interactions, a likely interaction for the N-terminal peptide between R5 and Y93, with a percent occupancy of 100% (see **FIGURE 4B)**, is clearly lost for the central Aβ peptide in **FIGURE 4D**, **4E**, and **4F** as the possible corresponding interaction between K16 and Y93 has a 0% occupancy.

The above observations suggest that the FAED region of the alternate Aβ epitope still may form at least a few binding contacts with the gantenerumab antigen-combining site. This is in qualitative agreement with experimental observations for gantenerumab, which display an epitope in the VFFAEDVGSN region, but do not display an epitope emerging from neighboring regions of the central epitope sequence that include the sequence HHQKL for instance[23]. The much less stable binding free energy of 1.8 kcal/mol that we obtained for the central Aβ peptide bound to gantenerumab is also consistent with the observation that gantenerumab is unable to bind and alter soluble Aβ levels in contrast to what was observed for solanezumab, which preferentially recognizes the central Aβ epitope[23, 61].

The central Aβ peptide bound to crenezumab yielded a calculated binding free energy of -15.2 kcal/mol. The transposed N-terminal peptide had a weaker binding free energy of -3.1 kcal/mol. While the calculated free energy was smaller for the N-terminal peptide, the gap in free energy between the bound N-terminal and central Aβ peptides was much smaller for crenezumab than it was for gantenerumab. It is noteworthy to mention that the most stable transposed N-terminal peptide was obtained when the phenylalanine residue in the reverse N-terminal epitope, SDHRFEAD, was aligned with F20 in the original central Aβ epitope, HHQKLVFFAEDV, prior to model building. Binding in this reverse sense in the antigen-combining site of an anti-Aβ antibody has been noted before as solanezumab binds to the central Aβ peptide in this reverse sense compared to the orientation adopted by the N-terminal peptide in gantenerumab.[59] The original binding pose of the central Aβ peptide bound to crenezumab is shown in **FIGURE 5A** and **5B**. This may be compared to the binding pose observed for the N-terminal peptide bound to crenezumab during the production portion of our MD simulation in **FIGURE 5C**, **5D**, **5E**, and **5F**. Compared to the initial binding pose of the central Aβ peptide bound to crenezumab, it is seen that the binding site of crenezumab has had to distort considerably to bind to the N-terminal Aβ peptide in an alternate conformation.

The pattern in potential binding interactions for crenezumab is qualitatively similar to what we saw before with gantenerumab, but in general the percent occupancy for the various sites that we examined are usually higher for the transposed Aβ peptide bound to crenezumab. For comparison with **TABLE S1**, the results of percent occupancy calculations for crenezumab are available in **TABLE S2**. The percent occupancy is 99.0% for F19 to V94 when the central Aβ peptide is bound and is 100% for F4 to V94 when the N-terminal Aβ peptide is bound. The negatively charged E3 residue also appears to be attracted to a nearby binding pocket as seen in **FIGURE 5E** and **5F**; the percent occupancy for E22 to N52 is 100% when the central A$\beta$ peptide is bound and only drops to 92.28% for E3 to N52 when the N-terminal peptide is bound. In addition, the D23 to N53 percent occupancy of 36.94% when the central A$\beta$ peptide is bound has actually increased to 86.3% for D1 to N53 when the N-terminal A$\beta$ peptide is bound. Also, similar to the situation for gantenerumab, the positively charged residue in the alternate peptide does not stably bind to crenezumab. The percent occupancy of 100% for K16 to D101 when the central A$\beta$ peptide is bound drops to 0% for R5 to D101 when the N-terminal A$\beta$ peptide is bound.

In an attempt to understand why we were able to obtain a slightly stable binding free energy for the reverse epitope bound to crenezumab, it is noteworthy to point out the electrostatic similarities between the N-terminal DAEFRHD epitope and the reverse sequence of the central epitope KLVFFAED, which when written out in reverse becomes DEAFFVLK. When both sequences are compared in this way, it becomes apparent that the first four amino acids in the N-terminal sequence and the first four amino acids in the reverse central epitope sequence are very similar to each other, with DAEF and DEAF differing only by switching the internal positions of the alanine and glutamate residues. However, despite this similarity, the inability to interchange the binding of key arginine and lysine residues elsewhere in the sequence may help explain how crenezumab achieves its preference for the central A$\beta$ peptide. The inability for lysine and arginine residues to cross bind for both gantenerumab and crenezumab is likely due to a difference in the positioning of these residues in both the sequence and three-dimensional space; the arginine residue appears next to the DAEF sequence in the N-terminal epitope (DAEFRHDS) whereas the lysine residue appears three residues away from the corresponding FAED sequence in the central epitope (HHQKLVFFAEDV).

## Improving the binding affinity of PFA1 to pE3-A$\beta_{3-8}$

With an eye towards the future, we wanted to see if we could use a visual inspection of our MD simulations to rationally plan out single amino acid mutations that improve the calculated MMPBSA binding free energy between an anti-A$\beta$ antibody and a polymorphic A$\beta$ species. Such a computational approach has been successfully demonstrated before[62], and it is a cost-effective way to probe for antibody mutations that could potentially improve the binding strength and specificity of an antibody for an additional A$\beta$ target. The most promising mutants that are identified may then be produced and tested in a laboratory to confirm the predicted improvement in binding affinity at a later point.

In order to illustrate this approach, we examined a question posed by Gardberg et al. in their study of pE3-A$\beta_{3-8}$ bound to the anti-protofibril antibody, PFA1. In that study, the pE3-A$\beta_{3-8}$ amyloid peptide was shown to bind to PFA1 with less affinity ($K_D = 3000$ nM) than the wild type A$\beta_{2-7}$ peptide ($K_D = 60$ nM)[31]. pE3-A$\beta_{3-8}$ is still considered to be dangerous due to its prominent presence in Alzheimer's plaques, and a mutant antibody that can bind to both A$\beta$ species with high affinity would be more desirable as a potential drug candidate. To probe for such a mutant, we first ran a preliminary MD simulation of PFA1 bound to pE3-A$\beta_{3-8}$ and compared it to a MD simulation of PFA1 bound to A$\beta_{1-8}$ that had been used previously in our method validation. Of particular interest in the MD simulation for A$\beta_{1-8}$ was that the glutamate residue was found to be localized fairly well inside of its binding pocket as would be expected from analyzing the initial structure (see **FIGURE 3**). In contrast, the MD simulation for

pE3-A$\beta_{3-8}$ revealed that the terminal pyroglutamate (PCA3) residue was engaged in a tug-of-war of sorts between the glutamate binding pocket and another nearby binding pocket that normally belongs to either alanine or aspartate.

Based on the MD simulations, we proposed that the difference in the observed binding affinity was caused by either: 1) the PCA3 residue in pE3-A$\beta_{3-8}$ lacking the full negative charge of the glutamate residue, weakening its attraction to the glutamate binding pocket, or 2) the loss of alanine in pE3-A$\beta_{3-8}$, opening up the possibility for PCA3 to be attracted away from the glutamate binding pocket towards the other nearby unoccupied binding pocket. On these grounds, we introduced two sets of single amino acid substitution mutations into the binding pocket of PFA1, carried out MD simulations, and calculated the MMPBSA binding free energies between the mutant structures and the pE3-A$\beta_{3-8}$ peptide. The first two mutants, Y59A and N60A on the H chain, were designed to weaken the attraction of PCA3 towards the other nearby binding pocket. The next two mutants, S92K and H93K on the L chain, were designed to strengthen the positive electrostatic character in the glutamate binding pocket.

The MMPBSA results are given in **TABLE 6**, convergence plots for the calculated MMPBSA binding free energies are shown in **FIGURE S5**, and the results of percent occupancy calculations are available in **TABLE S3**. Two out of the four proposed mutants were able to lower the calculated binding free energy by an appreciable amount indicating that these mutations stabilized the bound structure. The calculated binding free energy for pE3-A$\beta_{3-8}$ bound to the N60A PFA1 mutant was -10.2 kcal/mol, and the calculated binding free energy for the Y59A PFA1 mutant was -8.4 kcal/mol. These were both more favorable than the binding free energy of -3.9 kcal/mol calculated for pE3-A$\beta_{3-8}$ bound to the PFA1 wildtype structure.

The more favorable binding free energies were also comparable to values we obtained of -14.3 kcal/mol and -10.4 kcal/mol for the original wild type A$\beta_{1-8}$ and A$\beta_{2-7}$ peptides bound to PFA1 respectively.

Examining the MD trajectories (**FIGURE 6**) revealed that our N60A mutation worked out more or less as we expected. The PCA3 residue was now localized closer to the two histidine residues near the glutamate binding pocket for the majority of the simulation time. For the N60A mutant, the percent occupancy for PCA3 to H27D of 97.9% and for PCA3 to H93 of 99.9% were large increases over the corresponding wildtype values of 19.7% and 44.0% respectively. In addition, the percent occupancy for PCA3 to S58, a residue near the N60A mutation site, was reduced from 50.9% in the wildtype to 8.6% in the N60A mutant.

In a somewhat similar way, our Y59A mutant appeared to stabilize PCA3 by reducing its movement, but this time the PCA3 residue localized more towards the other nearby binding pocket on the central right side as indicated by an increase in percent occupancy from 50.9% to 81.0% for PCA3 to S58. Unlike in the N60A mutant, the PCA3 residue showed only a modest increase in its localization towards the H27D and H93 residues with percent occupancy values of 25.8% and 62.4% compared to 19.7% and 44.0% in the wildtype respectively.

The other two mutations did not improve the binding affinity of pE3-A$\beta_{3-8}$ towards the mutant antibody. The S92K mutant had a calculated binding free energy of 5.3 kcal/mol that indicated a strong destabilization of the bound structure. In the S92K mutant we note that the PCA3 to H27D percent occupancy decreased considerably from 19.7% to 1.0%, and also the PCA3 to L96 percent occupancy increased from 17.3% to 76.1%, indicating that perhaps an increase in percent occupancy at this particular site is disruptive to binding.

The H93K mutant was also destabilized as its calculated binding free energy of -2.7 kcal/mol was a bit lower than that of the wild type pE3-A$\beta_{3-8}$ peptide bound to PFA1. It failed to pull PCA3 residues towards the H27D and H93 locations as we intended; the PCA3 to H27D percent occupancy was lowered from 19.7% in the wildtype to 0% in the H93K mutant while the PCA3 to K93 (the position which was previously H93) was lowered from 44.0% in the wildtype to 9.8% in the H93K mutant. Analysis of the MD trajectory also revealed that the pE3-A$\beta_{3-8}$ peptide had undergone drastic changes in its binding pose for the H93K mutant. The entire backbone of pE3-A$\beta_{3-8}$ was flipped outwards toward the solvent, which allowed the PCA3 residue to move into the deep central binding pocket, displacing phenylalanine. The PCA3 to L96 percent occupancy increased to 100% for the H93K mutant compared to 17.3% in the wildtype.

Given the rather large changes in binding pose and affinity for the pE3-A$\beta_{3-8}$ peptide towards our PFA1 mutants, it is also reasonable to ask what changes might take place for the binding of the original wild type A$\beta_{1-8}$ and A$\beta_{2-7}$ peptides to our two successful mutant antibodies. This is an important issue if we wish to find a single antibody that is capable of binding with high affinity to multiple A$\beta$ species. We therefore also examined MD simulations and calculated MMPBSA binding free energies for A$\beta_{1-8}$ and A$\beta_{2-7}$ bound to our N60A and Y59A mutant forms of PFA1. The binding free energies are given in **TABLE 6**, representative snapshots from the MD trajectories are shown in **FIGURE 7**, and the results of percent occupancy calculations for both wild type and mutant forms are available in **TABLE S4** and **S5** for A$\beta_{1-8}$ and A$\beta_{2-7}$ respectively.

For both mutants, the binding affinity for A$\beta_{1-8}$ was actually improved over the wildtype PFA1 antibody. The calculated binding free energy for A$\beta_{1-8}$ bound to the N60A PFA1 mutant was -18.0 kcal/mol, and the calculated binding free energy for the Y59A PFA1 mutant was -16.7 kcal/mol. One possible explanation for such an increase in binding affinity is that the percent occupancy of D1 to N27 increased to 99.9% in the N60A mutant and to 83.8% in the Y59A mutant compared to 79.1% in the wildtype. In contrast, both mutations slightly destabilized the binding of the A$\beta_{2-7}$ peptide as the calculated binding free energy for A$\beta_{2-7}$ bound to the N60A PFA1 mutant was -8.3 kcal/mol, and the calculated binding free energy for the Y59A PFA1 mutant was -9.7 kcal/mol. A plausible explanation is that the mutation interfered with the binding of alanine, which is present in A$\beta_{2-7}$, and also that a compensating D1 to N27 interaction is completely missing for this case.

Nevertheless, if we compare the computed binding free energies of the wild type PFA1 for A$\beta_{1-8}$, A$\beta_{2-7}$, and pE3-A$\beta_{3-8}$ species to that of our N60A and Y59A mutants, we see that we have generally improved the affinity for our various A$\beta$ species albeit with a modest sacrifice in affinity for A$\beta_{2-7}$. In practice, a compromise between the binding of A$\beta$ species for a given antibody can be circumvented entirely by going beyond the single antibody approach. Instead, a cocktail of similar antibodies can be used to target various key A$\beta$ species in a treatment regime. In this case for instance, both wildtype and N60A mutant forms could be used together in a proposed treatment option to maximize effectiveness.

## Discussion

Anti-A$\beta$ antibodies are currently the most advanced treatment option on the horizon for patients suffering from AD. The recent positive clinical results reported for adacanumab indicate the potential for these drugs to be effective at clearing plaque burden and reducing cognitive decline. The main issues at present are the serious autoimmune side effects associated with the stronger binding anti-A$\beta$ antibodies and the presence of important polymorphic forms of A$\beta$ that may not bind to an antibody drug candidate with the same high affinity as its primary A$\beta$ target, decreasing its effectiveness.

Although it remains a possibility to obtain new antibody drug candidates from standard drug screening procedures, these are expensive and time-consuming endeavors to undertake, and there is no way of knowing what impact the drug will have on a human population until a very late stage of drug development. Alternatively, there exists the possibility to rationally modify and improve current anti-A$\beta$ antibody drugs that have already undergone clinical trials and whose strengths and weaknesses as a drug candidate are at least somewhat understood. Here, we have outlined a computational approach for studying the antigen-combining site of anti-A$\beta$ antibodies using fragment-based docking and full molecular dynamics simulations accompanied by MMPBSA binding free energy calculations. Computational methods are a cost-effective way to study the binding properties of anti-A$\beta$ antibodies whose crystal structures have previously been made available for analysis.

Employing a fragment-based docking method provided us with a means to dock single amino acid residues in an unbiased fashion to the surface of anti-A$\beta$ antibodies to probe for key anchoring residues that are involved in the initial binding interaction. Phenylalanine emerged as a dominant interaction, displaying the most stable binding free energy and very consistently docking into the antigen-combining sites of all ten antibodies. This observation that phenylalanine is a central anchor in the binding of A$\beta$ peptides to anti-A$\beta$ antibodies is corroborated by experimental observations from various crystallographers[23, 26, 59-60].

Our docking approach borrows heavily from current ideas being used in computational fragment-based drug design. In these methods, potential binding sites are located by probing the surface of a large protein receptor using small fragments of a ligand as opposed to attempting to dock the entire ligand[32-33]. A potential disadvantage to using small fragments to probe the receptor is a decrease in the binding selectivity for the ligand; additional binding sites other than the primary site of interest are often identified as potential binding sites[63]. On the other hand, conformational possibilities for a small fragment are much lower than for a larger ligand, and it has been shown that the leading hot spots identified using computational fragment-based methods correlate well with actual ligand binding sites[33]. In the present case, we were fortunate to be examining antibodies where the location of the ligand binding site was already known beforehand.

In addition, MD simulations were used to study questions posed by structural biologists in prior work on anti-A$\beta$ antibodies. In our study of PFA1, we showed that such simulations were capable of reproducing experimental observations as well as providing new insights into previously observed experimental results. We also demonstrated that MD simulations can be used as a tool to assist in exploratory research aimed at unravelling additional binding sites, to study mechanisms of binding, and to predict useful mutations to engineer into new anti-A$\beta$ antibody drug candidates.

The most noteworthy observation from our MD exploration of cross binding was that our data indicates that crenezumab may bind to both N-terminal and central A$\beta$ epitopes, although the binding to the N-terminal region is predicted to be much weaker. This topic is of interest because cross binding between these two epitopes may be related to aducanumab's reported ability to bind to both oligomeric and fibril A$\beta$ species, and this may be a factor in its more optimistic outlook as a drug candidate. Additionally, it has been observed experimentally that crenezumab has a puzzling ability to bind to both soluble monomers, using the central epitope of A$\beta$, and to insoluble amyloid plaques where it is believed that only the N-terminal epitope is readily accessible[25-26]. This behavior contrasts sharply with solanezumab which can only bind to soluble monomers using the central epitope of A$\beta$ and not to fibril structures via the N-terminal epitope.

To explain the cross binding to both fibril and oligomeric species observed for crenezumab, Ultsch et al. suggested that the A$\beta$ fibril species may have defects that expose the central A$\beta$ epitope along the fiber axis to allow crenezumab to bind to it in a few locations[26]. While this is certainly possible, the slightly stable MMPBSA binding affinity that we observed for binding the N-terminal peptide to crenezumab presents an alternative explanation; a weak binding affinity for the N-terminal epitope would give crenezumab a chance to bind to fibril structures to some extent. Although crenezumab's binding affinity for the fiber form would be lower than for the monomer or oligomer forms that bind via the central epitope, the high effective concentration of potential N-terminal binding sites along a fiber axis could still account for the sporadic fiber binding pattern observed by Ultsch et al.[26] We should note that other possibilities for cross binding to fibers exist than the two proposed above. Ma et al. has recently pointed out that crenezumab can recognize the A$\beta$ 13-16 epitope which may also be exposed in fibers to allow the antibody to bind[64]. It is important to note that our analysis assumed that a conserved interaction between the antibody and phenylalanine was maintained if cross binding occurred, but this assumption may not necessarily hold if A$\beta$ 13-16 is responsible for cross binding. An interesting feature of the A$\beta$ 13-16 epitope is that it contains two side by side histidine residues, implying that a strong pH dependence for fiber binding may help decide experimentally between this possibility and the others mentioned above.

Concerning our MD methodology, our analysis of gantenerumab and crenezumab is necessarily more limited and qualitative than our analysis of the PFA1 system. In contrast to our study of PFA1, we lacked a full set of experimental binding affinities that would have allowed us to validate our gantenerumab and crenezumab data and draw more quantitative conclusions for this portion of the study. In addition, while we have identified a few short sequences in the N-terminal and central A$\beta$ epitopes that may initiate a cross binding event, other important aspects that can affect binding selectivity, in particular the role of entropy in conformational selection for the full extended polypeptide sequence[59, 64-66], still needs to be quantified and studied in more detail. In our MD simulations, we observed a large amount of conformational flexibility in the both the bound A$\beta$ peptides and the amyloid antigen-combining sites of our various model systems. Such conformational movement made it difficult to obtain converged free energy values for our system, which required long 300 ns MD simulations and extensive 50 ns MMPBSA calculations to obtain acceptable convergence. We also used residue-to-residue percent occupancy values in our analysis, as opposed to more precise atom-to-atom calculations, due to the large amount of conformational flexibility that we observed. However, available methods that attempt to quantify these effects for protein-ligand interactions in the form of entropy calculations are unsatisfactory[42], and much effort is currently aimed at improving the accuracy and efficiency of such calculations[67-70]. In particular, the single trajectory MD/MMPBSA method that we have employed does not take into account any significant conformational changes that may occur between the bound and unbound states, and the multi-trajectory method that attempts to address this issue is known to have major issues with convergence[69]. Therefore, omitting entropy calculations is a standard practice when we apply the single trajectory MD/MMPBSA approach as we have done here and in past work[42, 71]. Nevertheless, we do believe that a truly deep understanding of binding selectivity will need to take into account entropic effects in a more quantitative manner, and finding a way to accurately quantify the entropic effects in a highly flexible system like this would be an interesting avenue for a future study.

Finally, we introduced rational mutations into the PFA1 antibody in an attempt to improve its binding affinity towards the pE3-A$\beta$ species of A$\beta$. Given the high conformational flexibility of both the A$\beta$ peptide and the antibody itself, predicting useful mutations from static crystal structures would be difficult, if not impossible, to do. On the other hand, visual inspection of the MD simulations of the amyloid-antibody complex allowed us to rationally identify potential mutation hot spots on the antibody surface with relative ease. Two out of the four mutants were shown to stabilize the binding of pE3-A$\beta_{3-8}$ to PFA1. Other possible mutations may exist that can impact the binding in a similar or even better way. If a computational pre-screening approach such as this is carried out, it should produce a list of potential hot spot mutations that can then be tested in the lab to confirm the predicted effectiveness. If the results from the lab correlate with the predicted computational results, an improved drug candidate with a greater potential to treat AD can subsequently be developed.

## Acknowledgements

We would like to thank Drs. Hartmut Luecke, Charles Glabe, and Hiromi Arai for helpful advice the led to the improvement of our computational docking study and whose prior work on anti-amyloid antibodies helped inspire this project. This work was supported by National Institute of Health/NIGMS (GM093040 & GM079383).

# References

1. Masters, C. L.; Simms, G.; Weinman, N. A.; Multhaup, G.; McDonald, B. L.; Beyreuther, K., Amyloid plaque core protein in Alzheimer disease and Down syndrome. *Proc Natl Acad Sci U S A* **1985,** *82* (12), 4245-9.
2. Iqbal, K.; Alonso Adel, C.; Chen, S.; Chohan, M. O.; El-Akkad, E.; Gong, C. X.; Khatoon, S.; Li, B.; Liu, F.; Rahman, A.; Tanimukai, H.; Grundke-Iqbal, I., Tau pathology in Alzheimer disease and other tauopathies. *Biochim Biophys Acta* **2005,** *1739* (2-3), 198-210.
3. Dobson, C. M., Getting out of shape. *Nature* **2002,** *418* (6899), 729-30.
4. Selkoe, D. J.; Hardy, J., The amyloid hypothesis of Alzheimer's disease at 25 years. *EMBO Mol Med* **2016,** *8* (6), 595-608.
5. Weiner, H. L.; Frenkel, D., Immunology and immunotherapy of Alzheimer's disease. *Nat Rev Immunol* **2006,** *6* (5), 404-16.
6. Pul, R.; Dodel, R.; Stangel, M., Antibody-based therapy in Alzheimer's disease. *Expert Opin Biol Ther* **2011,** *11* (3), 343-57.
7. Mavoungou C; Schindowski K; Atta-ur-Rahman (Ed.), *Immunotherapy with Anti-Aβ Monoclonal Antibodies in Alzhiemer's Disease: A Critical Review on the Molecules in the Pipelines with Regulatory Considerations*. Bentham Books: 2013; Vol. 1, p 3-85.
8. Orgogozo, J. M.; Gilman, S.; Dartigues, J. F.; Laurent, B.; Puel, M.; Kirby, L. C.; Jouanny, P.; Dubois, B.; Eisner, L.; Flitman, S.; Michel, B. F.; Boada, M.; Frank, A.; Hock, C., Subacute meningoencephalitis in a subset of patients with AD after Abeta42 immunization. *Neurology* **2003,** *61* (1), 46-54.
9. Lemere, C. A.; Maier, M.; Peng, Y.; Jiang, L.; Seabrook, T. J., Novel Abeta immunogens: is shorter better? *Curr Alzheimer Res* **2007,** *4* (4), 427-36.
10. Salloway, S.; Sperling, R.; Gilman, S.; Fox, N. C.; Blennow, K.; Raskind, M.; Sabbagh, M.; Honig, L. S.; Doody, R.; van Dyck, C. H.; Mulnard, R.; Barakos, J.; Gregg, K. M.; Liu, E.; Lieberburg, I.; Schenk, D.; Black, R.; Grundman, M.; Bapineuzumab 201 Clinical Trial, I., A phase 2 multiple ascending dose trial of bapineuzumab in mild to moderate Alzheimer disease. *Neurology* **2009,** *73* (24), 2061-70.
11. Siemers, E. R.; Friedrich, S.; Dean, R. A.; Gonzales, C. R.; Farlow, M. R.; Paul, S. M.; Demattos, R. B., Safety and changes in plasma and cerebrospinal fluid amyloid beta after a single administration of an amyloid beta monoclonal antibody in subjects with Alzheimer disease. *Clin Neuropharmacol* **2010,** *33* (2), 67-73.
12. Farlow, M.; Arnold, S. E.; van Dyck, C. H.; Aisen, P. S.; Snider, B. J.; Porsteinsson, A. P.; Friedrich, S.; Dean, R. A.; Gonzales, C.; Sethuraman, G.; DeMattos, R. B.; Mohs, R.; Paul, S. M.; Siemers, E. R., Safety and biomarker effects of solanezumab in patients with Alzheimer's disease. *Alzheimers Dement* **2012,** *8* (4), 261-71.
13. Abbott, A.; Dolgin, E., Failed Alzheimer's trial does not kill leading theory of disease. *Nature* **2016,** *540* (7631), 15-16.
14. The Lancet, N., Solanezumab: too late in mild Alzheimer's disease? *Lancet Neurol* **2017,** *16* (2), 97.
15. Hawkes, N., Promise of new Alzheimer's drug is dashed after lack of evidence. *BMJ* **2016,** *355*, i6362.

16. Imbimbo, B. P.; Ottonello, S.; Frisardi, V.; Solfrizzi, V.; Greco, A.; Seripa, D.; Pilotto, A.; Panza, F., Solanezumab for the treatment of mild-to-moderate Alzheimer's disease. *Expert Rev Clin Immunol* **2012,** *8* (2), 135-49.
17. Finder, V. H.; Glockshuber, R., Amyloid-beta aggregation. *Neurodegener Dis* **2007,** *4* (1), 13-27.
18. Russo, C.; Violani, E.; Salis, S.; Venezia, V.; Dolcini, V.; Damonte, G.; Benatti, U.; D'Arrigo, C.; Patrone, E.; Carlo, P.; Schettini, G., Pyroglutamate-modified amyloid beta-peptides--AbetaN3(pE)--strongly affect cultured neuron and astrocyte survival. *J Neurochem* **2002,** *82* (6), 1480-9.
19. Saido, T. C.; Iwatsubo, T.; Mann, D. M.; Shimada, H.; Ihara, Y.; Kawashima, S., Dominant and differential deposition of distinct beta-amyloid peptide species, A beta N3(pE), in senile plaques. *Neuron* **1995,** *14* (2), 457-66.
20. Cynis, H.; Frost, J. L.; Crehan, H.; Lemere, C. A., Immunotherapy targeting pyroglutamate-3 Abeta: prospects and challenges. *Mol Neurodegener* **2016,** *11* (1), 48.
21. Demattos, R. B.; Lu, J.; Tang, Y.; Racke, M. M.; Delong, C. A.; Tzaferis, J. A.; Hole, J. T.; Forster, B. M.; McDonnell, P. C.; Liu, F.; Kinley, R. D.; Jordan, W. H.; Hutton, M. L., A plaque-specific antibody clears existing beta-amyloid plaques in Alzheimer's disease mice. *Neuron* **2012,** *76* (5), 908-20.
22. Frost, J. L.; Liu, B.; Rahfeld, J. U.; Kleinschmidt, M.; O'Nuallain, B.; Le, K. X.; Lues, I.; Caldarone, B. J.; Schilling, S.; Demuth, H. U.; Lemere, C. A., An anti-pyroglutamate-3 Abeta vaccine reduces plaques and improves cognition in APPswe/PS1DeltaE9 mice. *Neurobiol Aging* **2015,** *36* (12), 3187-3199.
23. Bohrmann, B.; Baumann, K.; Benz, J.; Gerber, F.; Huber, W.; Knoflach, F.; Messer, J.; Oroszlan, K.; Rauchenberger, R.; Richter, W. F.; Rothe, C.; Urban, M.; Bardroff, M.; Winter, M.; Nordstedt, C.; Loetscher, H., Gantenerumab: a novel human anti-Abeta antibody demonstrates sustained cerebral amyloid-beta binding and elicits cell-mediated removal of human amyloid-beta. *J Alzheimers Dis* **2012,** *28* (1), 49-69.
24. Muhs, A.; Hickman, D. T.; Pihlgren, M.; Chuard, N.; Giriens, V.; Meerschman, C.; van der Auwera, I.; van Leuven, F.; Sugawara, M.; Weingertner, M. C.; Bechinger, B.; Greferath, R.; Kolonko, N.; Nagel-Steger, L.; Riesner, D.; Brady, R. O.; Pfeifer, A.; Nicolau, C., Liposomal vaccines with conformation-specific amyloid peptide antigens define immune response and efficacy in APP transgenic mice. *Proc Natl Acad Sci U S A* **2007,** *104* (23), 9810-5.
25. Adolfsson, O.; Pihlgren, M.; Toni, N.; Varisco, Y.; Buccarello, A. L.; Antoniello, K.; Lohmann, S.; Piorkowska, K.; Gafner, V.; Atwal, J. K.; Maloney, J.; Chen, M.; Gogineni, A.; Weimer, R. M.; Mortensen, D. L.; Friesenhahn, M.; Ho, C.; Paul, R.; Pfeifer, A.; Muhs, A.; Watts, R. J., An effector-reduced anti-beta-amyloid (Abeta) antibody with unique abeta binding properties promotes neuroprotection and glial engulfment of Abeta. *J Neurosci* **2012,** *32* (28), 9677-89.
26. Ultsch, M.; Li, B.; Maurer, T.; Mathieu, M.; Adolfsson, O.; Muhs, A.; Pfeifer, A.; Pihlgren, M.; Bainbridge, T. W.; Reichelt, M.; Ernst, J. A.; Eigenbrot, C.; Fuh, G.; Atwal, J. K.; Watts, R. J.; Wang, W., Structure of Crenezumab Complex with Abeta Shows Loss of beta-Hairpin. *Sci Rep* **2016,** *6*, 39374.
27. Crespi, G. A.; Hermans, S. J.; Parker, M. W.; Miles, L. A., Molecular basis for mid-region amyloid-beta capture by leading Alzheimer's disease immunotherapies. *Sci Rep* **2015,** *5*, 9649.

28. Sevigny, J.; Chiao, P.; Bussière, T.; Weinreb, P. H.; Williams, L.; Maier, M.; Dunstan, R.; Salloway, S.; Chen, T.; Ling, Y.; O'Gorman, J.; Qian, F.; Arastu, M.; Li, M.; Chollate, S.; Brennan, M. S.; Quintero-Monzon, O.; Scannevin, R. H.; Arnold, H. M.; Engber, T.; Rhodes, K.; Ferrero, J.; Hang, Y.; Mikulskis, A.; Grimm, J.; Hock, C.; Nitsch, R. M.; Sandrock, A., The antibody aducanumab reduces Aβ plaques in Alzheimer's disease. *Nature* **2016,** *537* (7618), 50-6.
29. Kastanenka, K. V.; Bussiere, T.; Shakerdge, N.; Qian, F.; Weinreb, P. H.; Rhodes, K.; Bacskai, B. J., Immunotherapy with Aducanumab Restores Calcium Homeostasis in Tg2576 Mice. *J Neurosci* **2016,** *36* (50), 12549-12558.
30. Case, D.A.; Betz, R.M.; Botello-Smith, W.; Cerutti, D.S.; Cheatham III, T.E.; Darden, T.A.; Duke, R.E.; Giese, T.J.; Gohlke, H.; Goetz, A.W.; Homeyer, N.; Izadi, S.; Janowski, P.; Kaus, J.; Kovalenko, A.; Lee, T.S.; LeGrand, S.; Li, P.; Lin, C.; Luchko, T.; Luo, R.; Madej, B.; Mermelstein, D.; Merz, K.M.; Monard, G.; Nguyen, H.; Nguyen, H.T.; Omelyan, I.; Onufriev, A.; Roe, D.R.; Roitberg, A.; Sagui, C.; Simmerling, C.L.; Swails, J.; Walker, R.C.; Wang, J.; Wolf, R.M.; Wu, X.; Xiao, L.; York, D.M.; Kollman, P.A., AMBER 2016, University of California, San Francisco.
31. Gardberg, A.; Dice, L.; Pridgen, K.; Ko, J.; Patterson, P.; Ou, S.; Wetzel, R.; Dealwis, C., Structures of Abeta-related peptide--monoclonal antibody complexes. *Biochemistry* **2009,** *48* (23), 5210-7.
32. Brenke, R.; Kozakov, D.; Chuang, G. Y.; Beglov, D.; Hall, D.; Landon, M. R.; Mattos, C.; Vajda, S., Fragment-based identification of druggable 'hot spots' of proteins using Fourier domain correlation techniques. *Bioinformatics* **2009,** *25* (5), 621-7.
33. Kozakov, D.; Hall, D. R.; Jehle, S.; Luo, L.; Ochiana, S. O.; Jones, E. V.; Pollastri, M.; Allen, K. N.; Whitty, A.; Vajda, S., Ligand deconstruction: Why some fragment binding positions are conserved and others are not. *Proc Natl Acad Sci U S A* **2015,** *112* (20), E2585-94.
34. MacCallum, J. L.; Bennett, W. F.; Tieleman, D. P., Distribution of amino acids in a lipid bilayer from computer simulations. *Biophys J* **2008,** *94* (9), 3393-404.
35. O'Boyle, N. M.; Banck, M.; James, C. A.; Morley, C.; Vandermeersch, T.; Hutchison, G. R., Open Babel: An open chemical toolbox. *J Cheminform* **2011,** *3*, 33.
36. Koes, D. R.; Baumgartner, M. P.; Camacho, C. J., Lessons learned in empirical scoring with smina from the CSAR 2011 benchmarking exercise. *J Chem Inf Model* **2013,** *53* (8), 1893-904.
37. Webb, B.; Sali, A., Comparative Protein Structure Modeling Using MODELLER. *Curr Protoc Bioinformatics* **2016,** *54*, 5 6 1-5 6 37.
38. Pettersen, E. F.; Goddard, T. D.; Huang, C. C.; Couch, G. S.; Greenblatt, D. M.; Meng, E. C.; Ferrin, T. E., UCSF Chimera--a visualization system for exploratory research and analysis. *J Comput Chem* **2004,** *25* (13), 1605-12.
39. Srinivasan, J.; Cheatham, T. E.; Cieplak, P.; Kollman, P. A.; Case, D. A., Continuum solvent studies of the stability of DNA, RNA, and phosphoramidate - DNA helices. *Journal of the American Chemical Society* **1998,** *120* (37), 9401-9409.
40. Kollman, P. A.; Massova, I.; Reyes, C.; Kuhn, B.; Huo, S. H.; Chong, L.; Lee, M.; Lee, T.; Duan, Y.; Wang, W.; Donini, O.; Cieplak, P.; Srinivasan, J.; Case, D. A.; Cheatham, T. E., Calculating structures and free energies of complex molecules: Combining molecular mechanics and continuum models. *Accounts Chem Res* **2000,** *33* (12), 889-897.

41.   Gohlke, H.; Case, D. A., Converging free energy estimates: MM-PB(GB)SA studies on the protein-protein complex Ras-Raf. *Journal of Computational Chemistry* **2004,** *25* (2), 238-250.
42.   Yang, T. Y.; Wu, J. C.; Yan, C. L.; Wang, Y. F.; Luo, R.; Gonzales, M. B.; Dalby, K. N.; Ren, P. Y., Virtual screening using molecular simulations. *Proteins-Structure Function and Bioinformatics* **2011,** *79* (6), 1940-1951.
43.   Miller, B. R.; McGee, T. D.; Swails, J. M.; Homeyer, N.; Gohlke, H.; Roitberg, A. E., MMPBSA.py: An Efficient Program for End-State Free Energy Calculations. *Journal of Chemical Theory and Computation* **2012,** *8* (9), 3314-3321.
44.   Wang, C. H.; Nguyen, P. H.; Pham, K.; Huynh, D.; Le, T. B. N.; Wang, H. L.; Ren, P. Y.; Luo, R., Calculating protein-ligand binding affinities with MMPBSA: Method and error analysis. *Journal of Computational Chemistry* **2016,** *37* (27), 2436-2446.
45.   Luo, R.; Moult, J.; Gilson, M. K., Dielectric screening treatment of electrostatic solvation. *Journal of Physical Chemistry B* **1997,** *101* (51), 11226-11236.
46.   Wang, J.; Cai, Q.; Li, Z.-L.; Zhao, H.-K.; Luo, R., Achieving energy conservation in Poisson-Boltzmann molecular dynamics: Accuracy and precision with finite-difference algorithms. *Chemical Physics Letters* **2009,** *468* (4-6), 112-118.
47.   Wang, J.; Luo, R., Assessment of Linear Finite-Difference Poisson-Boltzmann Solvers. *Journal of Computational Chemistry* **2010,** *31* (8), 1689-1698.
48.   Cai, Q.; Hsieh, M.-J.; Wang, J.; Luo, R., Performance of Nonlinear Finite-Difference Poisson-Boltzmann Solvers. *Journal of Chemical Theory and Computation* **2010,** *6* (1), 203-211.
49.   Wang, J.; Cai, Q.; Xiang, Y.; Luo, R., Reducing Grid Dependence in Finite-Difference Poisson-Boltzmann Calculations. *Journal of Chemical Theory and Computation* **2012,** *8* (8), 2741-2751.
50.   Botello-Smith, W. M.; Luo, R., Applications of MMPBSA to Membrane Proteins I: Efficient Numerical Solutions of Periodic Poisson-Boltzmann Equation. *J Chem Inf Model* **2015,** *55* (10), 2187-99.
51.   Cai, Q.; Wang, J.; Zhao, H.-K.; Luo, R., On removal of charge singularity in Poisson-Boltzmann equation. *Journal of Chemical Physics* **2009,** *130* (14).
52.   Lwin, T. Z.; Luo, R., Overcoming entropic barrier with coupled sampling at dual resolutions. *Journal of Chemical Physics* **2005,** *123* (19).
53.   Lwin, T. Z.; Zhou, R. H.; Luo, R., Is Poisson-Boltzmann theory insufficient for protein folding simulations? *Journal of Chemical Physics* **2006,** *124* (3).
54.   Case, D. A.; Cheatham, T. E.; Darden, T.; Gohlke, H.; Luo, R.; Merz, K. M.; Onufriev, A.; Simmerling, C.; Wang, B.; Woods, R. J., The Amber biomolecular simulation programs. *J Comput Chem* **2005,** *26* (16), 1668-88.
55.   Tan, C.; Tan, Y.-H.; Luo, R., Implicit nonpolar solvent models. *Journal of Physical Chemistry B* **2007,** *111* (42), 12263-12274.
56.   Cai, Q.; Ye, X.; Wang, J.; Luo, R., On-the-Fly Numerical Surface Integration for Finite-Difference Poisson-Boltzmann Methods. *Journal of Chemical Theory and Computation* **2011,** *7* (11), 3608-3619.
57.   Olsson, N.; Wallin, S.; James, P.; Borrebaeck, C. A.; Wingren, C., Epitope-specificity of recombinant antibodies reveals promiscuous peptide-binding properties. *Protein Sci* **2012,** *21* (12), 1897-910.

58. DiBrino, M.; Parker, K. C.; Margulies, D. H.; Shiloach, J.; Turner, R. V.; Biddison, W. E.; Coligan, J. E., The HLA-B14 peptide binding site can accommodate peptides with different combinations of anchor residues. *J Biol Chem* **1994,** *269* (51), 32426-34.

59. Ma, B.; Zhao, J.; Nussinov, R., Conformational selection in amyloid-based immunotherapy: Survey of crystal structures of antibody-amyloid complexes. *Biochim Biophys Acta* **2016,** *1860* (11 Pt B), 2672-81.

60. Gardberg, A. S.; Dice, L. T.; Ou, S.; Rich, R. L.; Helmbrecht, E.; Ko, J.; Wetzel, R.; Myszka, D. G.; Patterson, P. H.; Dealwis, C., Molecular basis for passive immunotherapy of Alzheimer's disease. *Proc Natl Acad Sci U S A* **2007,** *104* (40), 15659-64.

61. DeMattos, R. B.; Bales, K. R.; Cummins, D. J.; Dodart, J. C.; Paul, S. M.; Holtzman, D. M., Peripheral anti-A beta antibody alters CNS and plasma A beta clearance and decreases brain A beta burden in a mouse model of Alzheimer's disease. *Proc Natl Acad Sci U S A* **2001,** *98* (15), 8850-5.

62. Clark, L. A.; Boriack-Sjodin, P. A.; Eldredge, J.; Fitch, C.; Friedman, B.; Hanf, K. J.; Jarpe, M.; Liparoto, S. F.; Li, Y.; Lugovskoy, A.; Miller, S.; Rushe, M.; Sherman, W.; Simon, K.; Van Vlijmen, H., Affinity enhancement of an in vivo matured therapeutic antibody using structure-based computational design. *Protein Sci* **2006,** *15* (5), 949-60.

63. Ludlow, R. F.; Verdonk, M. L.; Saini, H. K.; Tickle, I. J.; Jhoti, H., Detection of secondary binding sites in proteins using fragment screening. *Proc Natl Acad Sci U S A* **2015,** *112* (52), 15910-5.

64. Zhao, J.; Nussinov, R.; Ma, B., Mechanisms of recognition of amyloid-beta (Abeta) monomer, oligomer, and fibril by homologous antibodies. *J Biol Chem* **2017,** *292* (44), 18325-18343.

65. Nussinov, R.; Ma, B.; Tsai, C. J., Multiple conformational selection and induced fit events take place in allosteric propagation. *Biophys Chem* **2014,** *186*, 22-30.

66. Weikl, T. R.; Paul, F., Conformational selection in protein binding and function. *Protein Sci* **2014,** *23* (11), 1508-18.

67. Meirovitch, H.; Cheluvaraja, S.; White, R. P., Methods for calculating the entropy and free energy and their application to problems involving protein flexibility and ligand binding. *Curr Protein Pept Sci* **2009,** *10* (3), 229-43.

68. de Ruiter, A.; Oostenbrink, C., Free energy calculations of protein-ligand interactions. *Curr Opin Chem Biol* **2011,** *15* (4), 547-52.

69. Steinbrecher, T.; Labahn, A., Towards accurate free energy calculations in ligand protein-binding studies. *Curr Med Chem* **2010,** *17* (8), 767-85.

70. Lin, J. H., Accommodating protein flexibility for structure-based drug design. *Curr Top Med Chem* **2011,** *11* (2), 171-8.

71. Greene, D.; Botello-Smith, W. M.; Follmer, A.; Xiao, L.; Lambros, E.; Luo, R., Modeling Membrane Protein-Ligand Binding Interactions: The Human Purinergic Platelet Receptor. *J Phys Chem B* **2016,** *120* (48), 12293-12304.

# Tables

| Sequence |
|---|
| Aβ 1-23: DAEFRHDSGYEVHHQKLVFFAED |
| Aβ 24-42: VGSNKGAIIGLMVGGVVIA |

| RANK | PFA1 | PFA1 (apo) | PFA2 | PFA2 (apo) | BAPI | SOLA | GANT | CREN | CREN (apo) | PONE |
|---|---|---|---|---|---|---|---|---|---|---|
| 1 | -5.9(F) | -6.1(F) | -6.4(F) | -6.3(Y) | -6.2(Y) | -5.9(Y) | -6.2(Y) | -6.1(Y) | -5.8(F/Y) | -6.2(Y) |
| 2 | -5.8(Y) | -6.1(Y) | -5.9(Y) | -6.1(F) | -5.9(F) | -5.6(F) | -5.7(F) | -5.9(F) | -4.9(H) | -6.0(F) |
| 3 | -5.0(R) | -5.2(R) | -5.1(R) | -5.2(H) | -5.2(H) | -5.2(H) | -4.8(K) | -5.3(H) | -4.7(R/Q) | -5.3(H) |
| 4 | -4.8(Q) | -4.9(H) | -5.1(H) | -4.9(Q) | -5.1(E) | -5.0(R) | -4.7(I) | -5.2(R) | -4.6(L) | -4.8(D/E/L) |
| 5 | -4.6(H) | -4.8(E/N/Q) | -4.8(E/Q) | -4.8(E) | -5.0(D) | -4.7(K) | -4.6(Q/R) | -4.7(E/Q) | -4.5(E) | -4.7(I/K/Q) |
| 6 | -4.5(E/N) | -4.7(D) | -4.6(D/L) | -4.6(I) | -4.9(R) | -4.6(Q) | -4.5(D/E/H/L/V) | -4.6(N) | -4.4(N/K/I) | -4.5(V) |
| 7 | -4.4(D/L) | -4.6(I/L) | -4.5(I) | -4.5(L/N) | -4.8(N) | -4.5(E) | -4.0(M) | -4.5(L) | -4.3(D/V) | -4.2(R) |
| 8 | -4.2(I) | -4.2(V) | -4.3(K) | -4.4(D/K) | -4.7(I) | -4.4(I/L) | -3.9(N) | -4.4(V) | -4.1(M) | -4.0(M) |
| 9 | -3.9(V) | -4.1(K) | -4.1(V) | -4.3(R/V) | -4.6(L) | -4.1(N/V) | -3.7(S) | -4.3(D/I) | -4.0(S) | -3.9(S) |
| 10 | -3.8(M) | -3.9(S) | -4.0(M/S) | -4.2(M) | -4.5(K) | -3.9(D) | -3.4(A) | -4.2(M) | -3.6(A) | -3.6(A) |

**TABLE 1. The top binding affinities for single amino acid residues bound to the antigen-combining site of anti-Aβ antibodies.** Binding free energies are given in units of kcal/mol while the corresponding amino acid is indicated by using the standard single-letter amino acid code. All of the structures given are based on the holo crystal structure form of the antibody except for those marked "(apo)" which are based on the apo crystal structure. Antibody abbreviations: PFA1 = protofibril antibody 1, PFA2 = protofibril antibody 2, BAPI = bapineuzumab, SOLA = solanezumab, GANT = gantenerumab, CREN = crenezumab, PONE = ponezumab. The Aβ 1-42 amino acid sequence is also provided where residues in the N-terminal epitope are highlighted in red, residues in the central epitope are highlighted in blue, and residues that appear in both epitopes are highlighted in green.

# Tables

| Sequence |
|---|
| Aβ 1-23: DAEFRHDSGYEVHHQKLVFFAED |
| Aβ 24-42: VGSNKGAIIGLMVGGVVIA |

| # of Bound Structures | PFA1 | PFA1 (apo) | PFA2 | PFA2 (apo) | BAPI | SOLA | GANT | CREN | CREN (apo) | PONE |
|---|---|---|---|---|---|---|---|---|---|---|
| 9 | R | | F | | F/E/L/I/M/V | | | F/Q | F/K/M | |
| 8 | | | Y | H/N/K/I | Y/D/H | | F | R/H/K | R/H/Y/L | F/Y/I/L/K |
| 7 | | F | | H/L/M | Y/D/Q/L/R | N/K/Q | E/F/K | | E/L/I | E/I | |
| 6 | F/K/L | Y | R/K/I | F/E/M/V | R | I | I | Y/M | | H/D/V/N |
| 5 | Y | R/H | E | | A/S | Y/M | K | D/G/V/N | | Q/M |
| 4 | | E/Q/K/I | Q | A | | H/L/R/Q | L | A | Q | R |
| 3 | E/I/Q | D/L/N/M | | S | G | D/V | A/E/Q/Y | | D/A/N/V | A/E/S |
| 2 | V/M | V | D | G | | | G/N | D/M/V | S | |
| 1 | D/H/N | A/S | | G/V/N/S | | A/S | H/N/R/S | | S/G | G |
| 0 | A/G/S | G | A | | | | G | | | |

**TABLE 2. The number of docked structures found at the antigen-combining site of anti-Aβ antibodies.** The number of docked structures for each amino acid residue that were found in the antigen-combining site of the antibody are reported in the table above. Each amino acid is indicated by using the standard single-letter amino acid code. Antibody abbreviations: PFA1 = protofibril antibody 1, PFA2 = protofibril antibody 2, BAPI = bapineuzumab, SOLA = solanezumab, GANT = gantenerumab, CREN = crenezumab, PONE = ponezumab. The Aβ 1-42 amino acid sequence is also provided where residues in the N-terminal epitope are highlighted in red, residues in the central epitope are highlighted in blue, and residues that appear in both epitopes are highlighted in green.

# Tables

| Complex Structure | MMPBSA | Experimental |
|---|---|---|
| PFA1 (A$\beta_{1-8}$) | -14.3 | -10.2** |
| PFA1 (A$\beta_{2-7}$) | -10.9 | -9.7* |
| PFA1 (pE3-A$\beta_{3-8}$) | -3.9 | -7.6 |
| PFA1 (Grip1) | -6.7 | -7.5 |
| PFA1 (Ror2) | -17.7 | -10.5 |
| PFA2 (A$\beta_{1-8}$) | -14.0 | -10.4** |
| PFA2 (A$\beta_{2-7}$) | -9.5 | -9.1* |
| PFA2 (pE3-A$\beta_{3-8}$) | -4.0 | -6.7 |
| PFA2 (Grip1) | -1.4 | -6.9 |
| PFA2 (Ror2) | -10.8 | -9.7 |

**TABLE 3. Method validation of our MMPBSA calculations for A$\beta$ peptides bound to PFA1 and PFA2.** The results of MMPBSA binding free energy calculations and experimental values are given in units of kcal/mol. Experimental values were taken from Gardberg et al.[60] and were converted to free energy values as described in **Materials and Methods**. * denotes cases where the experimental binding affinity values were reported as a range, and we used the average value of that range as the experimental binding affinity. ** denotes the experimental value given for A$\beta_{1-40}$ binding to both PFA1 and PFA2. The standard error of the mean was 0.1 kcal/mol for each structure.

# Tables

| Aβ Peptide | MMPBSA | Experiment |
|---|---|---|
| A$\beta_{2-7}$ (AEFRHD) | -10.9 | -9.7 |
| Grip1 (AKFRHD) | -6.7 | -7.5 |
| Pos4 (AEIRHD) | 6.8 | No Binding |

**TABLE 4. The importance of phenylalanine in the binding of A$\beta_{2-7}$ to PFA1.** The results of MMPBSA binding free energy calculations, in units of kcal/mol, are given to demonstrate the severity of the phenylalanine Pos4 mutation (AEIRHD) in the binding of A$\beta_{2-7}$ to the antibody PFA1. The less severe Grip1 mutation (AKFRHD) was also included for comparison. The crystal structure of A$\beta_{2-8}$ bound to PFA1 (PDB ID: 2IPU) was used both for constructing the wildtype A$\beta_{2-7}$ peptide and for generating the homology models of the two other A$\beta$ peptide mutants. Experimental values were taken from Gardberg et al.[31, 60] and were converted to free energy values as described in **Materials and Methods**. The standard error of the mean was 0.1 kcal/mol for each structure.

## Tables

| Complex Structure | MMPBSA | Experimental |
|---|---|---|
| gantenerumab (DAEFRHDSGYE) | -33.8 | -10.7 |
| gantenerumab forward (HHQKLVFFAEDV) | 1.8 | - |
| crenezumab (EVHHQKLVFFAEDVG) | -15.2 | -11.7 |
| crenezumab reverse 2 (SDHRFEAD) | -3.1 | - |

**TABLE 5. MMPBSA binding free energy results for N-terminal and central A$\beta$ peptides bound to gantenerumab and crenezumab.** The crystal structures of N-terminal A$\beta$ (DAEFRHDSGYE) bound to gantenerumab (PDB ID: 5CSZ) and central A$\beta$ (EVHHQKLVFFAEDVG) bound to crenezumab (PDB ID: 5VZY) were used both to calculate the MMPBSA binding free energy for the crystal structure-based complexes and for generating the four homology models of the transposed epitopes for each. Only the most stable of the four homology models of the transposed epitopes for each antibody were used for analysis. These are listed as gantenerumab forward and crenezumab reverse 2. The other structures were less stable and were omitted from our analysis (data not shown). The colored type for phenylalanine reveals the residue that was used to line up the phenylalanine residue in each homology model to the phenylalanine residues(s) in the original crystal structure (residues that are colored the same indicate the phenylalanine residue that was used for the alignment). Experimental binding free energies are given for the antibody structures bound to the monomer form of A$\beta$[23, 26]. The results of MMPBSA binding free energy calculations are given in units of kcal/mol. The standard error of the mean was 0.1 kcal/mol for each structure.

# Tables

| Complex Structure | MMPBSA |
|---|---|
| $A\beta_{1-8}$-PFA1 wildtype | -14.3 |
| $A\beta_{2-7}$-PFA1 wildtype | -10.9 |
| pE3-$A\beta_{3-8}$-PFA1 wildtype | -3.9 |
| pE3-$A\beta_{3-8}$-PFA1 mutant (Y59A (H chain)) | -8.4 |
| pE3-$A\beta_{3-8}$-PFA1 mutant (N60A (H chain)) | -10.2 |
| pE3-$A\beta_{3-8}$-PFA1 mutant (S92K (L chain)) | 5.3 |
| pE3-$A\beta_{3-8}$-PFA1 mutant (H93K (L chain)) | -2.7 |
| $A\beta_{1-8}$-PFA1 mutant (N60A (H chain)) | -18.0 |
| $A\beta_{2-7}$-PFA1 mutant (N60A (H chain)) | -8.3 |
| $A\beta_{1-8}$-PFA1 mutant (Y59A (H chain)) | -16.7 |
| $A\beta_{2-7}$-PFA1 mutant (Y59A (H chain)) | -9.7 |

**TABLE 6. MMPBSA binding free energy results for $A\beta_{1-8}$, $A\beta_{2-7}$, and pE3-$A\beta_{3-8}$ bound to the wildtype and several mutant forms of PFA1.** The crystal structure of pE3-$A\beta_{3-8}$ bound to PFA1 (PDB ID: 3EYS) was used both for the wildtype pE3-$A\beta_{3-8}$-PFA1 complex and for generating the homology models of the four pE3-$A\beta_{3-8}$-PFA1 mutant complexes. The crystal structure of $A\beta_{2-8}$ (PDB ID: 2IPU) was used to generate the homology models of the $A\beta_{1-8}$ and $A\beta_{2-7}$ peptides bound to the N60A and Y59A mutant forms of PFA1. The results of MMPBSA binding free energy calculations are given in units of kcal/mol. The standard error of the mean was 0.1 kcal/mol for each structure.

# Figures

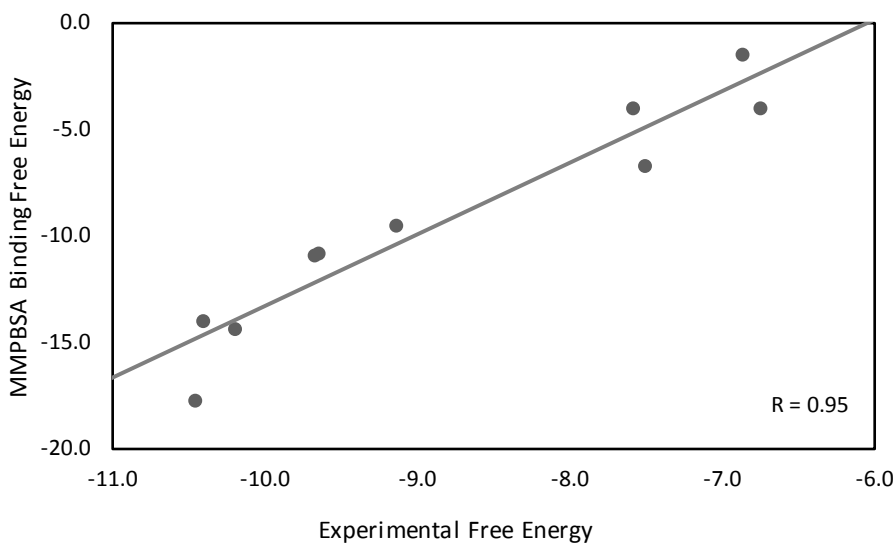

**FIGURE 1. Method validation for MMPBSA free energy calculations of various Aβ peptides bound to the antibodies PFA1 and PFA2.** The original PFA1, PFA2, and pE3-Aβ$_{3-8}$ crystal structures (PDB IDs: 2IPU, 2R0W, and 3EYS) served as the basis for constructing homology models for all the other bound peptide structures. Our calculated MMPBSA binding free energy values were compared to the experimental binding affinity values reported by Gardberg et al. for their entire data set.[60] All free energy values are given in units of kcal/mol.

# Figures

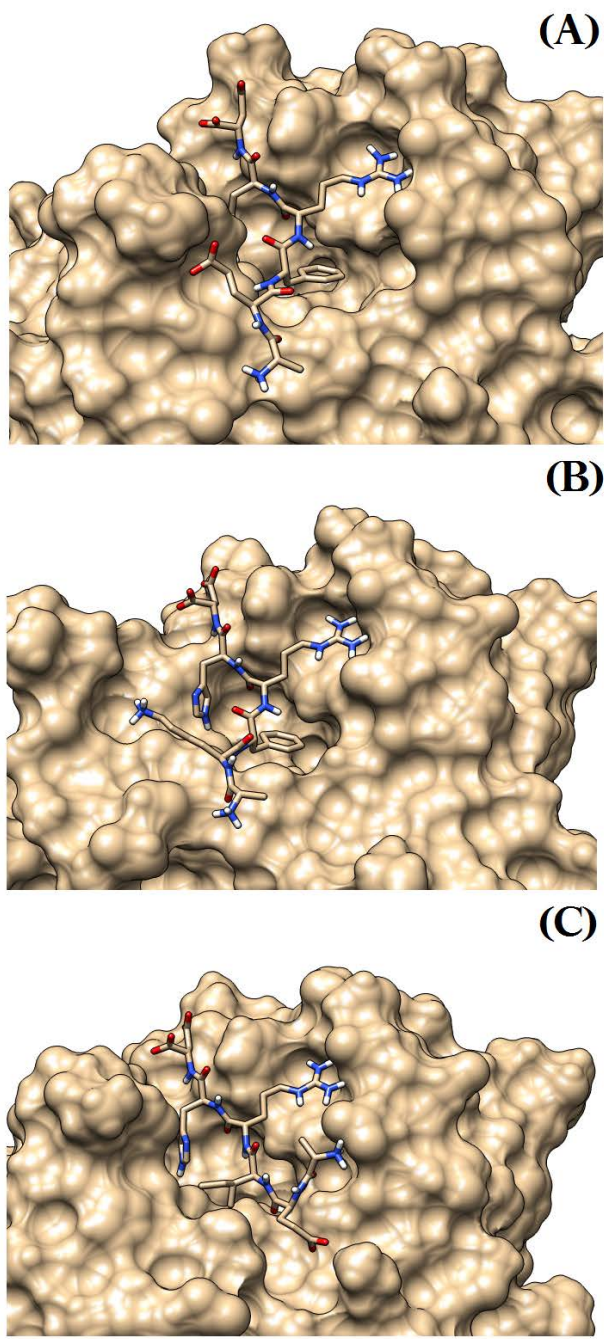

**FIGURE 2. Comparison of the binding pose for A$\beta_{2-7}$ peptide variants bound to PFA1.** Three A$\beta$ peptides are shown bound to PFA1: (A) A$\beta_{2-7}$, (B) Grip1, and (C) the Pos4 mutant. All three images were taken at the halfway point of the production portion of the MD simulation.

**Figures**

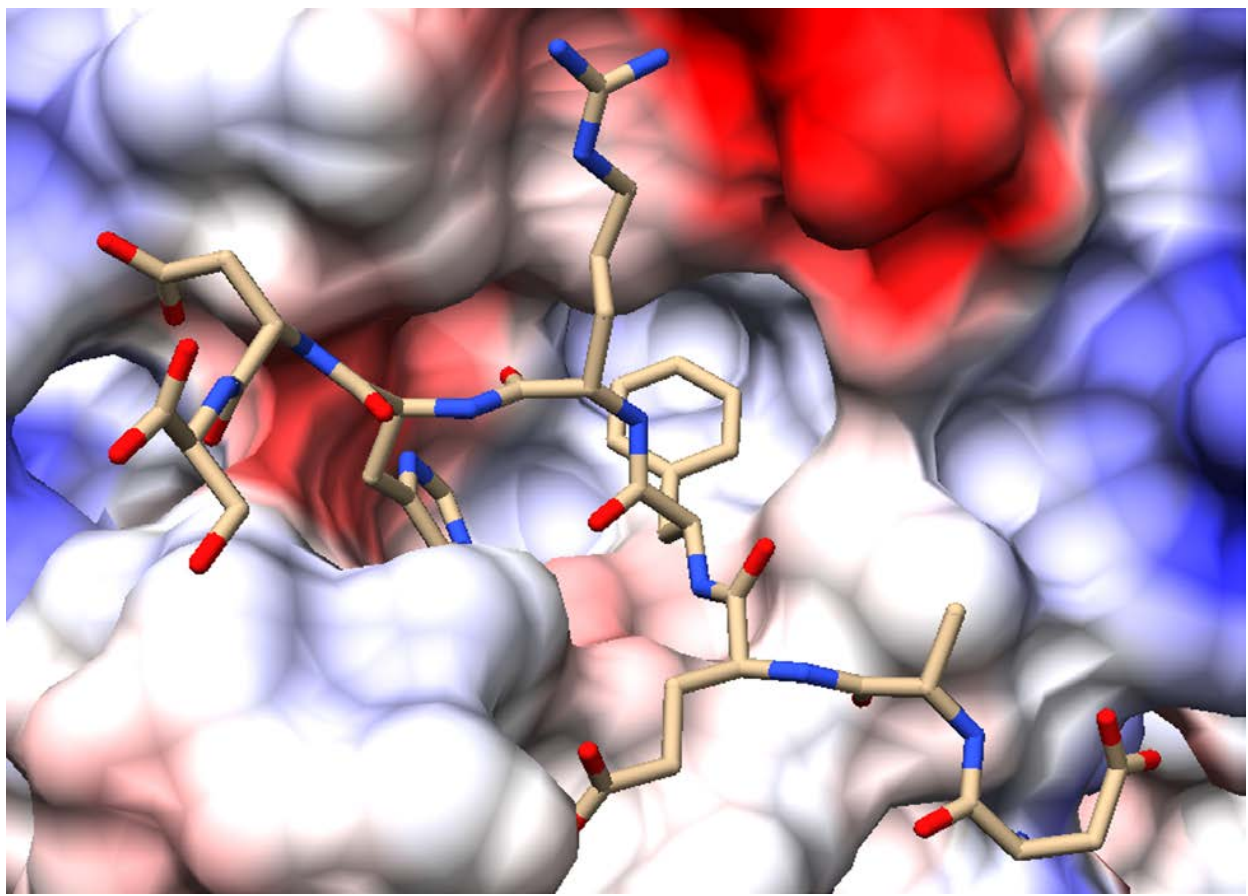

**FIGURE 3. Electrostatic contacts for A$\beta_{1-8}$ bound to PFA1.** A surface map of PFA1 is provided which shows the key electrostatic contacts made between the PFA1 antibody and charged residues in A$\beta_{1-8}$. The N-terminal aspartate residue (D1) on the A$\beta$ peptide appears in the lower right corner of the figure. Negatively charged regions are depicted in red while positively charged regions are shown in blue. Coulombic Surface Coloring was used to depict electrostatic contacts where the scale ranges from -10 kcal/mol*e (pure red, negative region) to 0 kcal/mol*e (pure white, neutral region) to 10 kcal/mol*e (pure blue, positive region).

# Figures

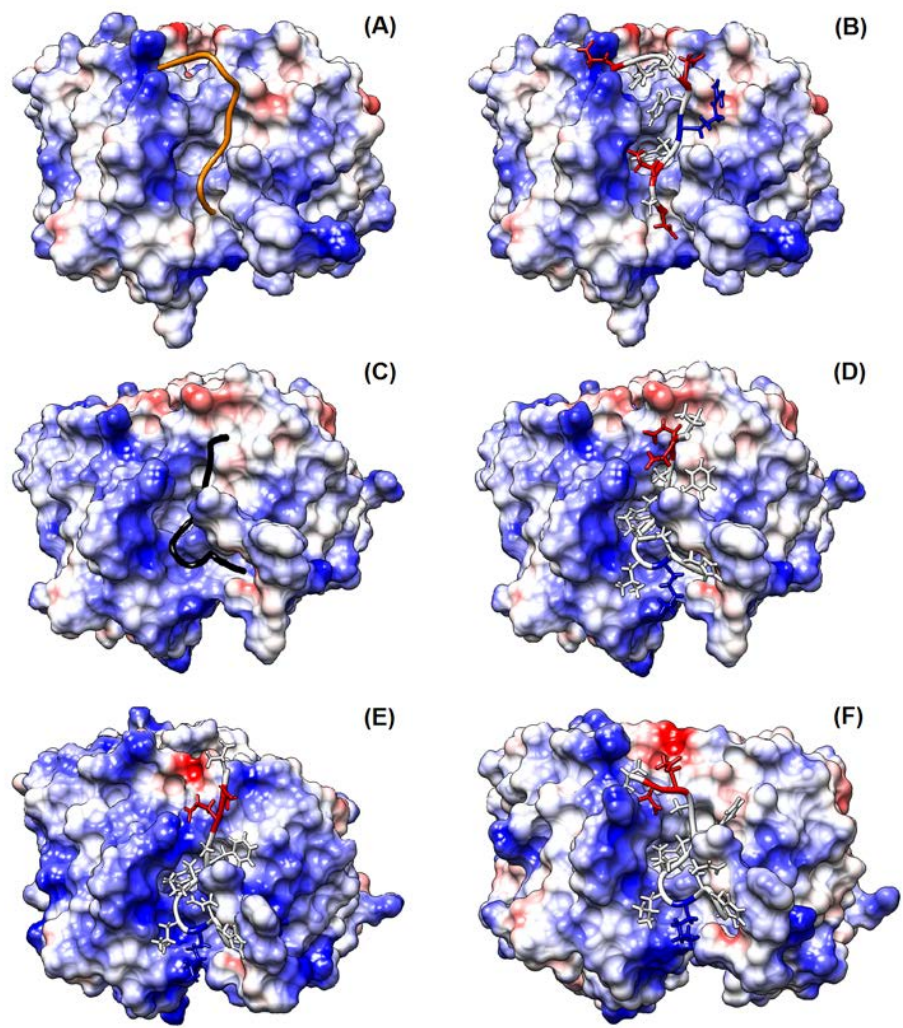

**FIGURE 4. Gantenerumab bound to both N-terminal and central Aβ peptides.** Gantenerumab is shown bound to the Aβ peptide containing the N-terminal epitope (PDB ID: 5CSZ) in the first frame of the MD simulation in (A) and (B). Structures (C), (D), (E), and (F) show the most stable central Aβ peptide bound to gantenerumab in the forward sequence (HHQKLVFFAEDV) across the gantenerumab antigen-combining site taken from the first frame (C and D), the middle frame (E), and the last frame (F) of the MD production run. In all structures, the N-terminus end of the peptide appears towards the bottom of the antigen-combining site while the C-terminus appears near the top. Coulombic Surface Coloring was used to depict electrostatic contacts on the antibody surface where the scale ranges from -10 kcal/mol*e (pure red, negative) to 0 kcal/mol*e (pure white, neutral) to 10 kcal/mol*e (pure blue, positive). The residues on the peptide were colored as acidic (red), basic (blue), or neutral (white).

# Figures

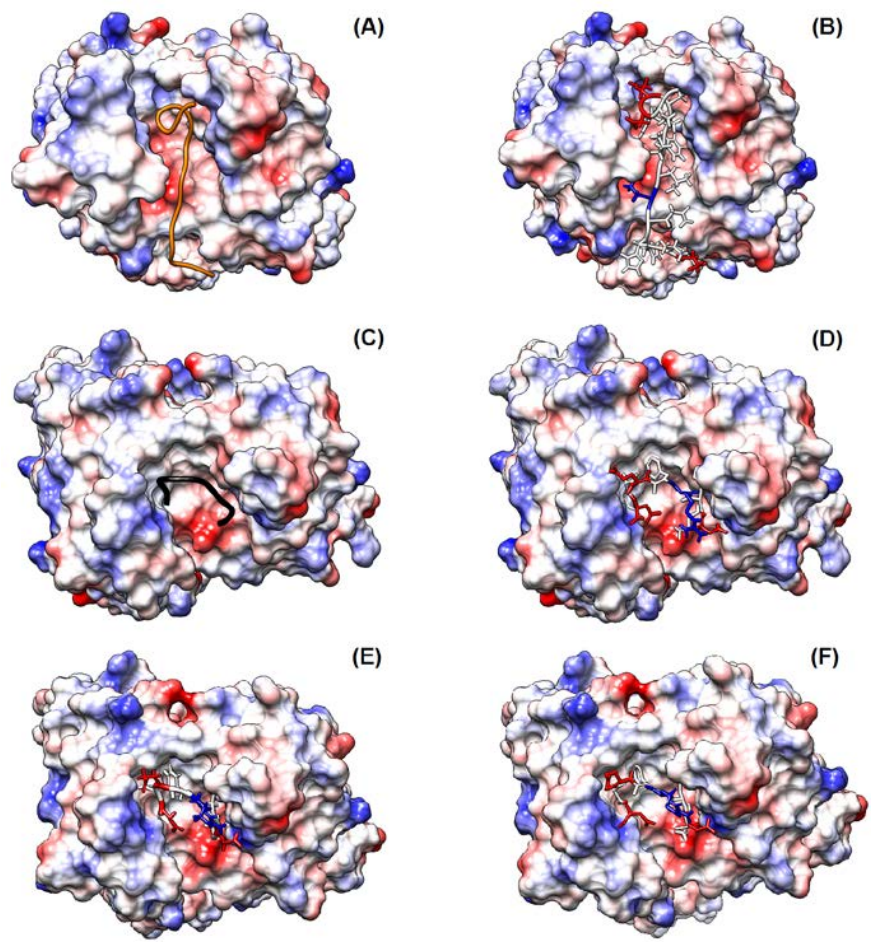

**FIGURE 5. Crenezumab bound to both central and N-terminal Aβ peptides.** Crenezumab is shown bound to the Aβ peptide containing the central epitope (PDB ID: 5VZY) in the first frame of the MD simulation in (A) and (B). Structures (C), (D), (E), and (F) show the N-terminal Aβ peptide bound to crenezumab in the reverse sense (SDHRFEAD) across the crenezumab antigen-combining site as observed in the first frame (C and D), the middle frame (E), and the last frame (F) of the MD production run. In structures (A) and (B), the N-terminus end of the peptide appears towards the bottom of the antigen-combining site while the C-terminus appears near the top. For (C), (D), (E), and (F), the N-terminus end appears towards the upper left of the antigen-combining site while the C-terminus end appears towards the lower right of the antigen-combining site. Coulombic Surface Coloring was used to depict electrostatic contacts on the antibody surface where the scale ranges from -10 kcal/mol*e (pure red, negative region) to 0 kcal/mol*e (pure white, neutral region) to 10 kcal/mol*e (pure blue, positive region). The residues on the peptide were colored as acidic (red), basic (blue), or neutral (white).

# Figures

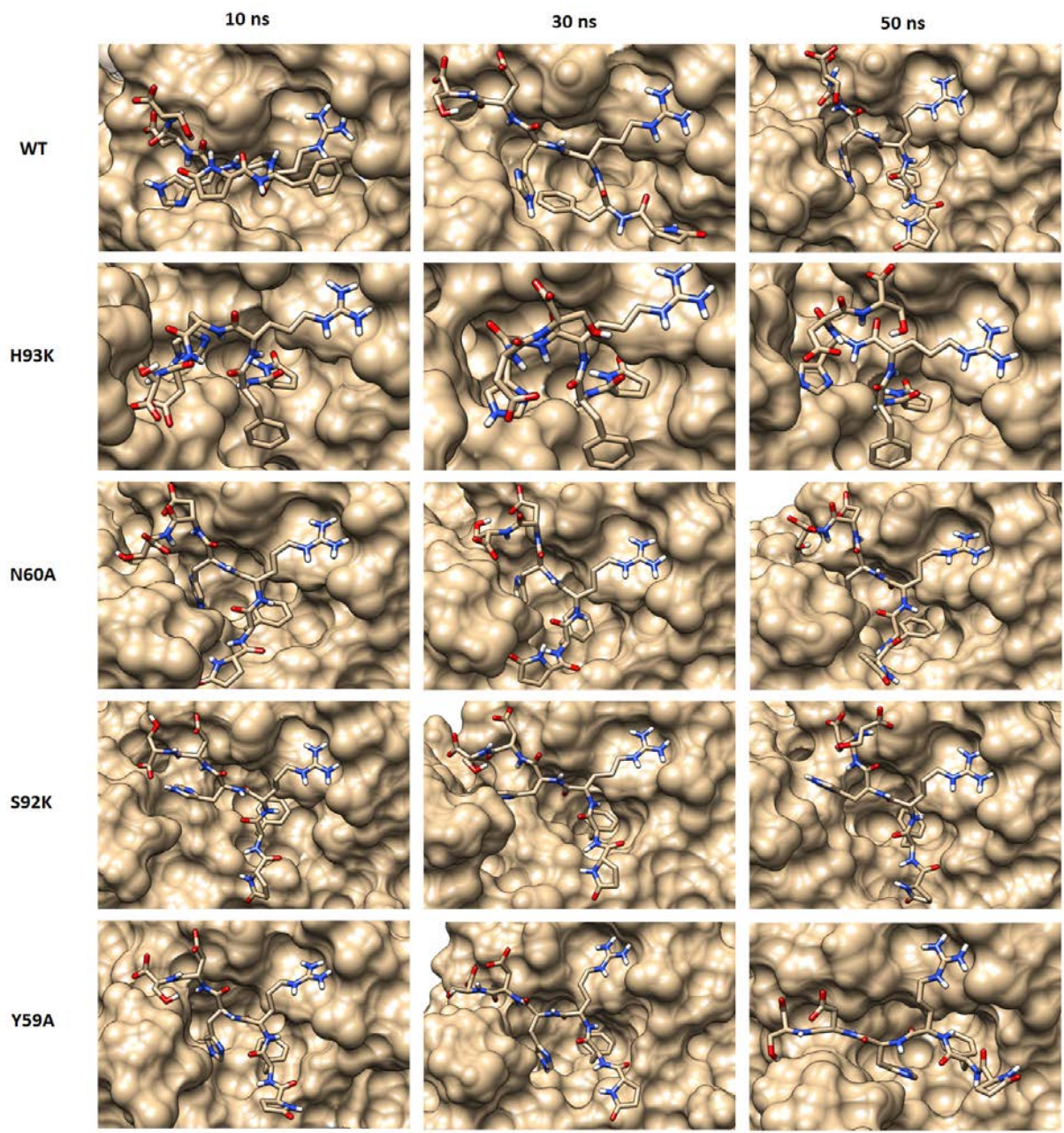

**FIGURE 6. Snapshots from the MD trajectories of pE3-A$\beta_{3-8}$ bound to wild type and mutant forms of PFA1.** The structures here are visualized as snapshots taken at 10 ns, 30 ns, and 50 ns during the production run of the MD simulation.

# Figures

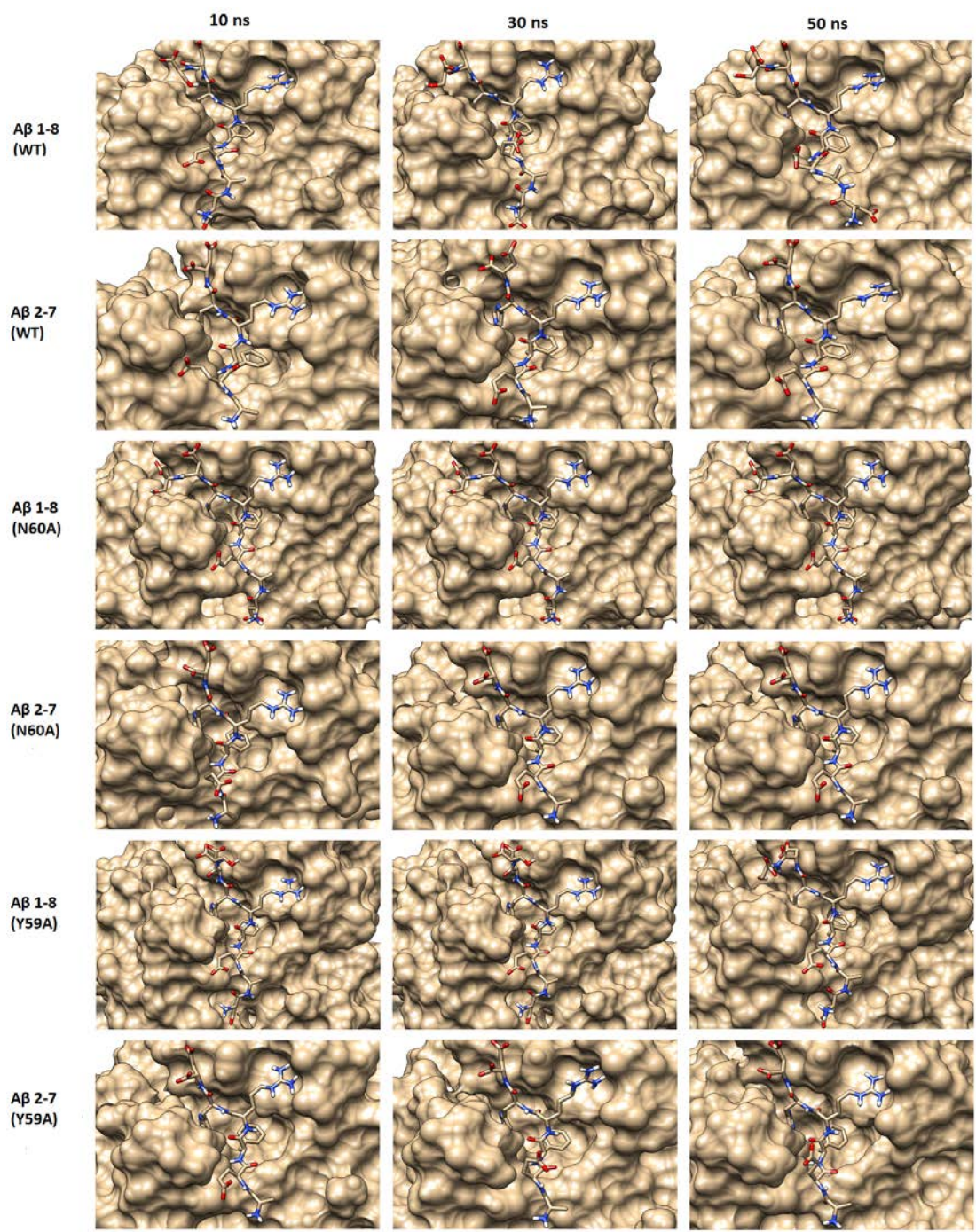

**FIGURE 7. Snapshots from the MD trajectories of A$\beta_{1-8}$ and A$\beta_{2-7}$ bound to wild type and mutant forms of PFA1.** The structures here are visualized as snapshots taken at 10 ns, 30 ns, and 50 ns during the production run of the MD simulation.